\documentclass[aps,prx,amsmath,twocolumn,amssymb,floatfix,showpacs,superscriptaddress,nofootinbib,longbibliography,10pt]{revtex4-2}
\usepackage[english]{babel}
\usepackage[utf8]{inputenc}
\usepackage[T1]{fontenc}
\usepackage{MnSymbol}
\usepackage{amsmath}
\usepackage{mathtools}
\usepackage{xcolor}
\usepackage{amsfonts}
\usepackage{amssymb}
\usepackage{braket}
\usepackage{float}
\usepackage{subfigure}
\usepackage{tikz}
\usepackage{multirow}
\usepackage[colorlinks=true,linktoc=page,linkcolor=blue,citecolor=blue,urlcolor=blue]{hyperref}
\mathchardef\mhyphen="2D % Define a "math hyphen"

\newcommand{\ie}{{i.e.,\,\,}}

\newcommand{\noi}{\noindent}

\newcommand{\non}{\nonumber}  
\usepackage[normalem]{ulem}
\definecolor{lime}{HTML}{A6CE39}
\usepackage{sidecap,tikz}
\DeclareRobustCommand{\orcidicon}{\hspace{-1.0mm}
	\begin{tikzpicture}
		\draw[lime, fill=lime] (0.0,0.0) 
		circle [radius=0.15] 
		node[white] {{\fontfamily{qag}\selectfont \tiny \,ID}};
		\draw[white, fill=white] (-0.0525,0.095) 
		circle [radius=0.007];
	\end{tikzpicture}
	\hspace{-3.0mm}
}

\newcommand{\orcidABS}{\href{https://orcid.org/0000-0002-4726-5247}{\orcidicon}}
\newcommand{\orcidAKG}{\href{https://orcid.org/0000-0003-0990-8341}{\orcidicon}}
\newcommand{\orcidPH}{\href{https://orcid.org/0000-0002-1866-2788}{\orcidicon}}
\newcommand{\orcidRSS}{\href{https://orcid.org/0000-0002-2978-3534}{\orcidicon}}
\newcommand{\orcidVAM}{\href{https://orcid.org/0000-0002-7838-0053}{\orcidicon}}

\allowdisplaybreaks

\begin{document}
%=============START of MAIN PAPER===============

\title{Quantum state transfer and maximal entanglement between distant qubits using a minimal quasicrystal pump}

\author{Arnob Kumar Ghosh\orcidAKG}
\email{arnob.ghosh@physics.uu.se}
\affiliation{Department of Physics and Astronomy, Uppsala University, Box 516, 75120 Uppsala, Sweden}

\author{Rubén Seoane Souto\orcidRSS}
\affiliation{Instituto de Ciencia de Materiales de Madrid (ICMM), Consejo Superior de Investigaciones Cient\'{i}ficas (CSIC), Sor Juana In\'{e}s de la Cruz 3, 28049 Madrid, Spain}

\author{Vahid Azimi-Mousolou\orcidVAM}
\affiliation{Department of Physics and Astronomy, Uppsala University, Box 516, 75120 Uppsala, Sweden}

\author{Annica M. Black-Schaffer\textsuperscript{\#}\orcidABS}
\email{annica.black-schaffer@physics.uu.se}
\affiliation{Department of Physics and Astronomy, Uppsala University, Box 516, 75120 Uppsala, Sweden}

\author{Patric Holmvall\textsuperscript{\#}\orcidPH}
\email{patric.holmvall@physics.uu.se}
\affiliation{Department of Physics and Astronomy, Uppsala University, Box 516, 75120 Uppsala, Sweden}

%----------------------------------------------------------------------
\begin{abstract}
Coherent quantum state transfer over macroscopic distances between non-neighboring elements in quantum circuits is a crucial component to increase connectivity and simplify quantum information processing. To facilitate such transfers, an efficient and easily controllable quantum pump would be highly beneficial. In this work, we demonstrate such a quantum pump based on a one-dimensional quasicrystal Fibonacci chain~(FC). In particular, we utilize the unique properties of quasicrystals to pump the edge-localized winding states between the two distant ends of the chain by only minimal manipulation of the FC at its end points. We establish the necessary conditions for successful state transfer within a fully time-dependent picture and also demonstrate robustness of the transfer protocol against disorder. We then couple external qubits to each end of the FC and establish highly adaptable functionality as a quantum bus with both on-demand switching of the qubit states and generation of maximally entangled Bell states between the qubits. Thanks to the minimal control parameters, the setup is well-suited for implementation across diverse experimental platforms, thus establishing quasicrystals as an efficient platform for versatile quantum information processing.
\end{abstract}
%----------------------------------------------------------------------

\maketitle

\begingroup
\renewcommand\thefootnote{\#}
\footnotetext{These authors contributed equally to this work.}
\endgroup

\setcounter{footnote}{0}

%====================================
\section{Introduction}
%====================================

Reliable transfer of quantum states and the distribution of entanglement over macroscopic distances~\cite{TrifunovicPRX2013,RochPRL2014} are central in emerging quantum technologies. This includes secure quantum communication and distributed quantum computing~\cite{BennettNature2000,DuanNature2001,GisinNC2007,NorthupNP2014,XuRMP2020,HuNRP2023}, underlying the development of the quantum internet, aimed at interconnecting quantum processors across extensive distances~\cite{KimbleNature2008,StephanieScience2018,AzumaRMP2023}.
Key to this endeavor are quantum buses capable of reversibly entangling and transferring quantum information between physical systems. Various physical platforms have already been proposed, including atoms coupled to cavities~\cite{CiracPRL1997,MatsukevichScience2004}, quantum spin chains~\cite{BosePRL2003,ChristandlPRL2004,YaoPRL2011}, and phononic systems~\cite{VermerschPRL2017}. However, these protocols often face challenges from disorder and noise, which degrade fidelity and hinder scalability.

A quantum pump based on protected boundary states of a topological material~\cite{hasan2010colloquium,qi2011topological,BernevigBOOK} can lead to topological quantum state transfer~\cite{CitroNRP2023}, which is naturally robust to disorder. To this end, quantum state transfer has already been proposed in several different topological systems~\cite{KrausPumpingPRL2012,YaoNatComm2013,VerbinPump2015,LangNQI2017,DlaskaQSI2017,MeiPRA2018,LonghiPRB2019,LonghiLandauZenerAQT2019,ZhengLiNaPRA2020,QiLuPRA2020,DAngelisPRR2020,PalaiodimopoulosPRA2021,CaoPRA2021,YuanAPLPh2021,QiPRRTR2021,WangPRA2022,ZhengLiNaYiPRApp2022,MiyazawaCommunMater2022,WeijiePRA2022,WangDaWeiPRA2023,ZhaonppjQuantumInf2023,Zurita2023fastquantumtransfer,RomeroPRApp2024,HanJinXuanPRApp2024,TianPRB2024,Fernandez2024,WangDaWeiPRA2024,ZhengCJP2025}. For example, there exist theory proposals for quantum state transfer based on the chiral edge state in quantum spin liquids~\cite{YaoNatComm2013,DlaskaQSI2017}, but facing substantial experimental challenges. A much larger set of work has instead focused on topological quantum state transfer in the dimerized Su–Schrieffer–Heeger~(SSH) model~\cite{LangNQI2017,MeiPRA2018,LonghiPRB2019,LonghiLandauZenerAQT2019,ZhengLiNaPRA2020,QiLuPRA2020,DAngelisPRR2020,PalaiodimopoulosPRA2021,CaoPRA2021,YuanAPLPh2021,QiPRRTR2021,WangPRA2022,ZhengLiNaYiPRApp2022,WeijiePRA2022,WangDaWeiPRA2023,ZhaonppjQuantumInf2023,Zurita2023fastquantumtransfer,RomeroPRApp2024,HanJinXuanPRApp2024,TianPRB2024,Fernandez2024,WangDaWeiPRA2024,ZhengCJP2025}, which is the simplest one-dimensional~(1D) topological setup and a periodic system. However, state transfer in an SSH chain is only achieved by dynamically switching the bond strength between intracell and intercell sites throughout the whole system~\cite{LangNQI2017,LonghiPRB2019,LonghiLandauZenerAQT2019,ZhengLiNaPRA2020,QiLuPRA2020,DAngelisPRR2020,PalaiodimopoulosPRA2021,CaoPRA2021,YuanAPLPh2021,QiPRRTR2021,WangPRA2022,ZhengLiNaYiPRApp2022,WeijiePRA2022,WangDaWeiPRA2023,ZhaonppjQuantumInf2023,Zurita2023fastquantumtransfer,RomeroPRApp2024,HanJinXuanPRApp2024,TianPRB2024,Fernandez2024,WangDaWeiPRA2024,ZhengCJP2025}. This leads to the number of operations to achieve a state transfer growing linearly with the distance, again limiting experimental feasibility. Here, we instead identify a 1D quasicrystal as a simpler and more efficient platform for achieving topological quantum state transfer, often not needing more than one or two bond switches in total, despite long distances.

Quasicrystals are quasiperiodic materials without translational symmetry but with exotic properties, such as discrete scale invariance and multifractal energy spectrum~\cite{KohmotoPRL1983,OstlundPRL1983,LevinePRL1984,ShechtmanPRL1984,KohmotoPRL1987,LuckPRB1989,GoldmanRMP1993,TanesePRL2014,ReisnerPRB2023,MoustajNJP2023,FrancaPRBL2024}. One of the most well-studied quasicrystals is the 1D Fibonacci chain~(FC)~\cite{JagannathanRMP2021,ZilberbergOME21}, which contains quasiperiodic modulation of either the hopping strength or onsite potential, following the Fibonacci sequence~\cite{JagannathanRMP2021,ZilberbergOME21}. Interestingly, the 1D FC can be associated with the topological properties of a two-dimensional~(2D) crystal Chern insulator under the inclusion of a so-called `phason' degree of freedom~\cite{KrausPRL2012,MadsenPRB2013,FlickerEPL2015,KrausNP2016,RaiPRB2021,JagannathanRMP2021,ZilberbergOME21,FanFP2022,jagannathanarXiv2025}. The resulting Chern numbers are associated with gap labels for each quasicrystal energy gap, which has been demonstrated in diffraction experiments~\cite{DareauPRL2017}. Subsequently, the FC supports topologically protected in-gap edge-localized states that  `wind' inside the quasicrystal gap, with a winding number equal to the gap label of that gap~\cite{JagannathanRMP2021,ZilberbergOME21}. These winding states have also been employed for quantum state transfer in photonic quasicrystals~\cite{KrausPumpingPRL2012,VerbinPump2015} and optical setups~\cite{SinghPRA2015}. However, previous work has not provided any protocol for on-demand transfer and, especially, demanded protocols that requires bond strengths to be switched throughout the whole length of the quasicrystal~\cite{KrausPumpingPRL2012,VerbinPump2015,SinghPRA2015}, which results in a similarly cumbersome setup as for the SSH chain~\cite{LangNQI2017,LonghiPRB2019,LonghiLandauZenerAQT2019,ZhengLiNaPRA2020,QiLuPRA2020,DAngelisPRR2020,PalaiodimopoulosPRA2021,CaoPRA2021,YuanAPLPh2021,QiPRRTR2021,WangPRA2022,ZhengLiNaYiPRApp2022,WeijiePRA2022,WangDaWeiPRA2023,ZhaonppjQuantumInf2023,Zurita2023fastquantumtransfer,RomeroPRApp2024,HanJinXuanPRApp2024,TianPRB2024,Fernandez2024,WangDaWeiPRA2024,ZhengCJP2025}. Finding a protocol that requires only minimal parameter switches for topological quantum state transfer, also over longer distances, is therefore absolutely crucial for any feasible and on-demand experimental realizations. Moreover, an important outstanding goal is to demonstrate that such a quantum pump can, in fact, be used for other important tasks such as entanglement generation and state preparation, paving the way to general-purpose quantum information processing.

\begin{figure}
    \centering
    \subfigure{\includegraphics[width=0.38\textwidth]{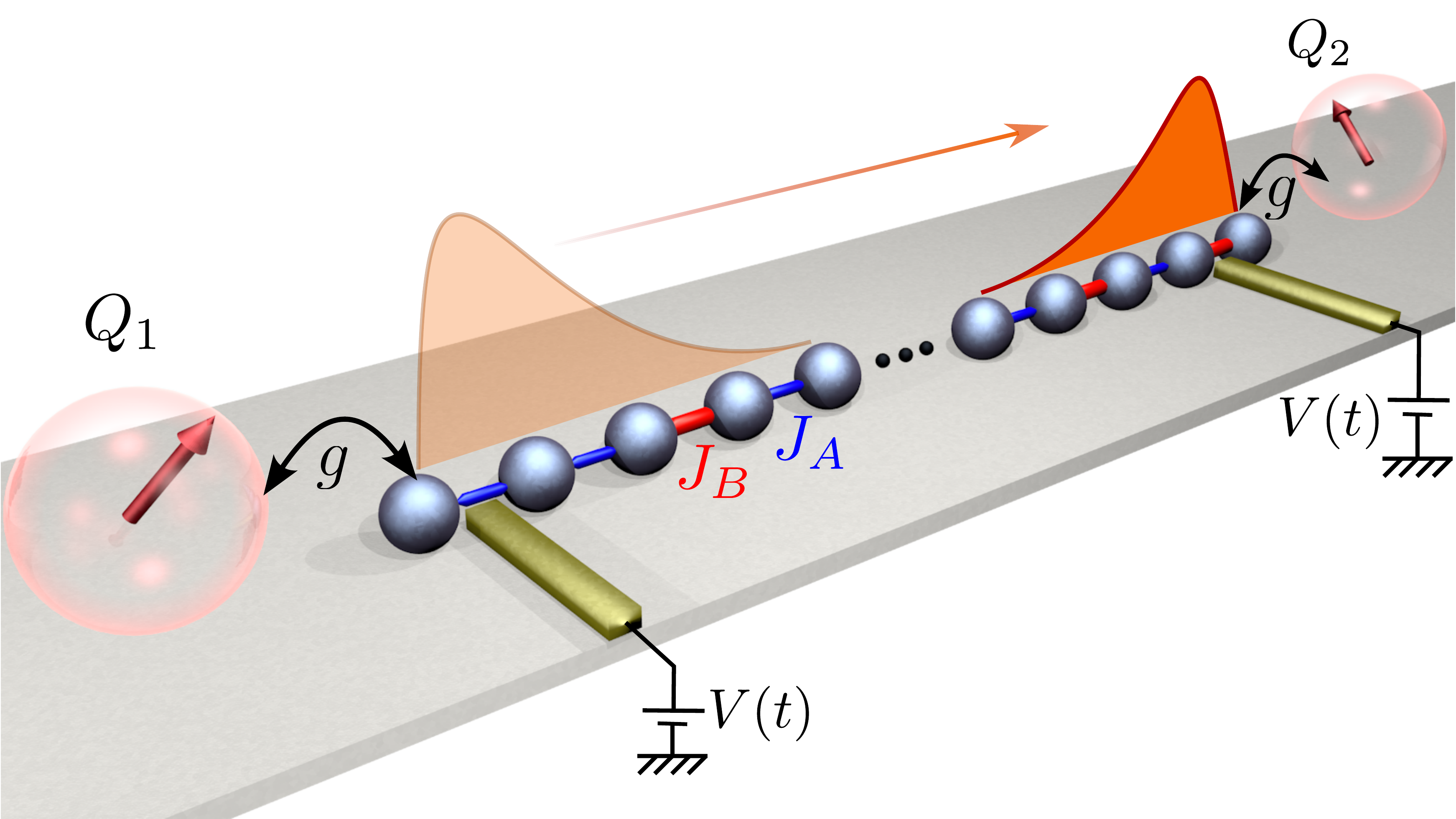}}
    \caption{The FQP consists of an FC with weak bonds $J_A$ (blue) and strong bonds $J_B$ (red). The transfer protocol $V(t)$ dynamically modifies the outermost bonds of the FC and induces a transfer of a winding state from the left end (light orange) to the right (dark orange). External qubits $Q_1$ and $Q_2$ are coupled to the FQP with coupling constant $g$. The FQP can transfer a state between the two qubits or generate maximally entangled states between them.
    }
    \label{Fig:Schematics}
\end{figure}

In this work, we exploit the intriguing quasiperiodic properties of the FC to develop a minimal topological Fibonacci quantum pump~(FQP), which requires only switching of one or two of the outermost bonds of the FC, see Fig.~\ref{Fig:Schematics} for an illustration. This dramatically reduces complexity and allows for simple and efficient implementation in different experimental frameworks, not just limited to optical setups~\cite{KrausPumpingPRL2012,VerbinPump2015,SinghPRA2015}, but also feasible with superconducting resonators~\cite{SplitthoffPRR2024}, Josephson junction arrays~\cite{johannsen2024fermionic}, or mechanical systems~\cite{MiyazawaCommunMater2022}. In particular, by studying the full-time evolution, we show how quantum state transfer is readily achieved with the FQP for a wide range of parameters and several simple transfer protocols. Importantly, we then utilize the FQP as a quantum bus to transmit information between two distant qubits connected at each end of the pump, see Fig.~\ref{Fig:Schematics}, and demonstrate both complete quantum state transfer and how to create maximally entangled Bell states between the qubits. Thus, the FQP may be controlled on demand for a variety of applications.

%~~~~~~~~~~~~~~~~~~~~~~~~~~~~~~~~~~~~~~~~~~~~~~~~~~~~~~~~~
%~~~~~~~~~~~~~~~~~~~~~~~~~~~~~~~~~~~~~~~~~~~~~~~~~~~~~~~~~
%~~~~~~~~~~~~~~~~~~~~~~~~~~~~~~~~~~~~~~~~~~~~~~~~~~~~~~~~~
%~~~~~~~~~~~~~~~~~~~~~~~~~~~~~~~~~~~~~~~~~~~~~~~~~~~~~~~~~

The remainder of the manuscript is organized as follows. In Sec.~\ref{Sec:Background}, we discuss the properties of the Fibonacci chain, which act as the building block of the pump. In Sec.~\ref{Sec:Pumping}, we demonstrate the pumping of the winding states in the FQP. In Sec.~\ref{Sec:StateTransfer}, we employ the FQP to transfer and create maximal entanglement between two distant qubits.  We conclude with a summary and discussion in Sec.~\ref{Sec:Summary}.

%====================================
\section{Background: Fibonacci chain} \label{Sec:Background}
%====================================

%--------------------------------------------
%\noi \textbf{Fibonacci chain}
%--------------------------------------------

\noi The 1D FC can be modeled by the Hamiltonian~\cite{JagannathanRMP2021}
\begin{align}
    H=\mu \sum_i c_i^\dagger c_i +   \sum_i J_i c_i^\dagger c_{i+1} + {\rm H.c.} \ ,
    \label{Eq:HamHopping}
\end{align}
where $\mu$ represents the chemical potential, $c_i^\dagger$ is the electronic creation operator at lattice site $i$, and $J_i$ is either a weak ($J_A$) or a strong ($J_B$) bond between two nearest neighbor sites, with $\left\{J_i\right \}$ belonging to the Fibonacci sequence $\left\{ C_n \right\}$. Here the $n$-th Fibonacci approximant $C_n$ is constructed through the concatenation rule $C_n=C_{n-1} \oplus C_{n-2}$ ($\forall n \geq 2$), with $C_0=J_B$ and $C_1=J_A$, leading to $C_2=J_A J_B$, $C_3=J_AJ_BJ_A$, $C_4=J_AJ_BJ_AJ_A J_B$, etc~\cite{JagannathanRMP2021}. Thus, the total number of bonds $N_B$ in $C_n$ is given by the Fibonacci number $F_n=F_{n-1} + F_{n-2}$, $\forall n \geq 2$, with $F_0=F_1=1$. Longer Fibonacci chains can also be constructed by repeating $l$ times the approximant $C_n$, such that the number of lattice sites is $L=l F_n+1$. We here always employ open boundary conditions for the chain, set $J_A=1$, $\mu=0$, and consider the dimensionless hopping ratio $\rho=J_B/J_A$. All energies (and times) are thus given in units of $J_A$. We verify that other parameters give qualitatively the same results. We also consider natural units, such that $c=\hbar=1$.

\begin{figure}
	\centering
	\subfigure{\includegraphics[width=0.49\textwidth]{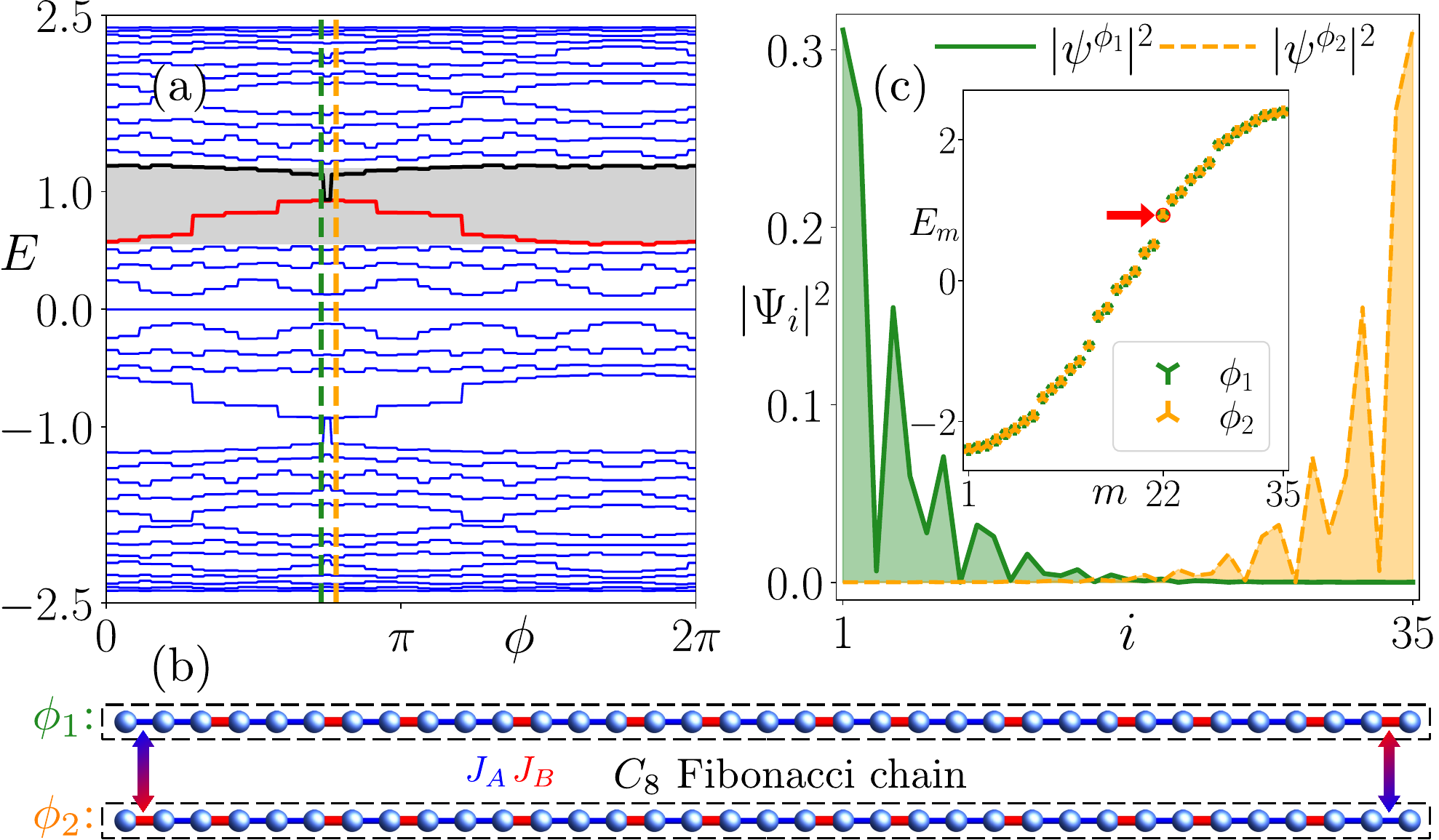}}
	\caption{(a) Eigenvalues $E$ as a function of phason angle $\phi$. Red and black curves represent winding states in one of the larger quasicrystal energy gaps (gray shaded). (b) Flipping between strong (red) and weak (blue) bonds indicated by arrows, when changing phason angle from $\phi_1$ (green) to $\phi_2$ (orange), also marked in (a). (c) Spatially resolved amplitude of the red winding state along the chain at $\phi_1$ (green) and $\phi_2$ (orange). Inset shows the eigenvalue spectrum $E_m$ as a function of the state index $m$ for phason angles $\phi_1$ (green) and $\phi_2$ (orange), with red dot and arrow marking the winding state. Here we use a $C_8$ FC with $L=F_8+1=35$ sites and $\rho=1.5$, $\phi_1=0.73 \pi$, and $\phi_2=0.78 \pi$.}
	\label{Fig:FQCbackground}
\end{figure}

The FC can alternatively be constructed using the sign of a characteristic function $\chi_i$, such that positive~(negative) values of $\chi_i$ set the $i$-th bond strength to $J_B~(J_A)$~\cite{JagannathanRMP2021}, with $\chi_i={\rm sgn}\left[\cos\left(2 \pi i \tau^{-1} + \pi \tau^{-1} + \phi\right) -\cos\left(\pi \tau^{-1}\right) \right]$, where $\tau= (1+ \sqrt{5})/2$ is the golden ratio and $\phi \in [0, 2 \pi)$ the phason angle. Changes in the phason angle account for pair-wise flipping of specific bond strengths~\cite{RontgenPRB2019} and as the phason angle $\phi$ varies from $0$ to $2\pi$, the topologically protected states `wind' across the energy gap~\cite{KrausPRL2012}. We illustrate this in Fig.~\ref{Fig:FQCbackground}(a) for a $C_8$ FC, where two in-gap states (red and black lines) wind once through one of the larger gaps (gray shaded) and are separated from the remaining states (blue lines). These states are therefore often referred to as winding states. The number of windings inside a specific energy gap is equal to its gap label, which can be connected to the Chern number of a parent 2D system using also the phasonic degree of freedom ~\cite{KrausPRL2012,MadsenPRB2013,FlickerEPL2015,KrausNP2016,JagannathanRMP2021,RaiPRB2021,ZilberbergOME21,FanFP2022,jagannathanarXiv2025}. The winding states with energies inside the gap are localized at one end of the FC, while they delocalize when approaching the gap edge~\cite{KrausPRL2012}. In Fig.~\ref{Fig:FQCbackground}(a), $\mathcal{C}=1$ for the highlighted winding states.

%~~~~~~~~~~~~~~~~~~~~~~~~~~~~~~~~~~~~~~~~~~~~~~~~~~~~~~~~~
%~~~~~~~~~~~~~~~~~~~~~~~~~~~~~~~~~~~~~~~~~~~~~~~~~~~~~~~~~
%~~~~~~~~~~~~~~~~~~~~~~~~~~~~~~~~~~~~~~~~~~~~~~~~~~~~~~~~~
%~~~~~~~~~~~~~~~~~~~~~~~~~~~~~~~~~~~~~~~~~~~~~~~~~~~~~~~~~

To construct an FQP with only small and easily implementable changes to the FC, we identify nearby values of the phason angle $\phi$ that amount to only flipping two outermost bonds in an FC approximant. This occurs, for example, for angles $\phi_1$ (green dashed line) and $\phi_2$ (orange dashed line) in Fig.~\ref{Fig:FQCbackground}(a), with the change of the outermost bonds indicated by arrows in Fig.~\ref{Fig:FQCbackground}(b). We then identify a large gap with a winding state well-separated from other states at these two phason angles as the state to pump, here the red winding state in Fig.~\ref{Fig:FQCbackground}(a). In Fig.~\ref{Fig:FQCbackground}(c) we plot the eigenstate of this winding state at phason angle $\phi_1$ (green) and find that it is well-localized at the left side of the FC, while for the $\phi_2$ angle (orange), its localization is completely switched to the right side. This change of localization of the winding state is the key component for using the FC as a quantum pump. 
Generally, such a pair of phason angles is always possible to identify and occurs for angles on either side of a critical angle $\phi_c$ where the winding state (red) approaches another winding state (black), see Fig.~\ref{Fig:FQCbackground}(a). 
The black winding state can also be employed as a pumping state, but with worse performance due to less edge-localization. Instead, the black state often becomes the lowest excited state during the transfer.
Moreover, all eigenvalues corresponding to the $\phi_1$ and $\phi_2$ FCs are unchanged since they are symmetrically oriented around $\phi_c$, see inset in Fig.~\ref{Fig:FQCbackground}(c).
Importantly, we only need to switch the two outermost bonds in order to go from phason angle $\phi_1$ to $\phi_2$ via $\phi_c$ and thereby obtain well-localized winding states at different ends of the chain, making it experimentally simpler than SSH-type topological pumps~\cite{LangNQI2017,LonghiPRB2019,LonghiLandauZenerAQT2019,ZhengLiNaPRA2020,QiLuPRA2020,DAngelisPRR2020,PalaiodimopoulosPRA2021,CaoPRA2021,YuanAPLPh2021,QiPRRTR2021,WangPRA2022,ZhengLiNaYiPRApp2022,WeijiePRA2022,WangDaWeiPRA2023,ZhaonppjQuantumInf2023,Zurita2023fastquantumtransfer,RomeroPRApp2024,HanJinXuanPRApp2024,TianPRB2024,Fernandez2024,WangDaWeiPRA2024,ZhengCJP2025} or earlier suggested FC pumps~\cite{KrausPumpingPRL2012,VerbinPump2015,SinghPRA2015}.

%====================================
\section{Minimal topological pump} \label{Sec:Pumping}
%====================================
%--------------------------------------------
% \vspace{2mm}
% \noi \textbf{Pumping winding states}
%--------------------------------------------

\noi Next, we demonstrate pumping of FC winding states. We focus on the $C_8$ FC in Fig.~\ref{Fig:Timedyn} in the main text, but other approximants of FCs show similar results, see Appendix~\ref{App:C6C7}. 
We consider the state to be initialized (at time $t=0)$ as the edge-localized winding state $\psi_w(0)=\psi^{\phi_1}_w$ at the left end of the chain for hopping sequence $\left\{ J_i^{\phi_1} \right\}$ given by phason angle $\phi_1$, see green state in Fig.~\ref{Fig:FQCbackground}(c).
We then apply an adiabatic one-step transfer protocol that switches only the two outermost bonds, marked by arrows in Fig.~\ref{Fig:FQCbackground}(b), such that we end up with the phason angle $\phi_2$ FC \footnote{For a short approximant, we may need to repeat the approximant more than once to obtain a desired system size, which then requires changing two bonds per repetition.}. The explicit time-dependence of the hopping is given in Eq.~\eqref{Eq:driveprotcol} in Appendix~\ref{App:TransferProtocol}, with adiabaticity controlled by the rate $\Omega$ and illustrated in the inset of Fig.~\ref{Fig:QubitCoupling}(a). We solve explicitly for the full time dependence to accurately capture the transfer, including non-adiabatic effects, see Appendix~\ref{App:TimeEvolution}.

%~~~~~~~~~~~~~~~~~~~~~~~~~~~~~~~~~~~~~~~~~~~~~~~~~~~~~~~~~
%~~~~~~~~~~~~~~~~~~~~~~~~~~~~~~~~~~~~~~~~~~~~~~~~~~~~~~~~~
\begin{figure}
	\centering
	\subfigure{\includegraphics[width=0.49\textwidth]{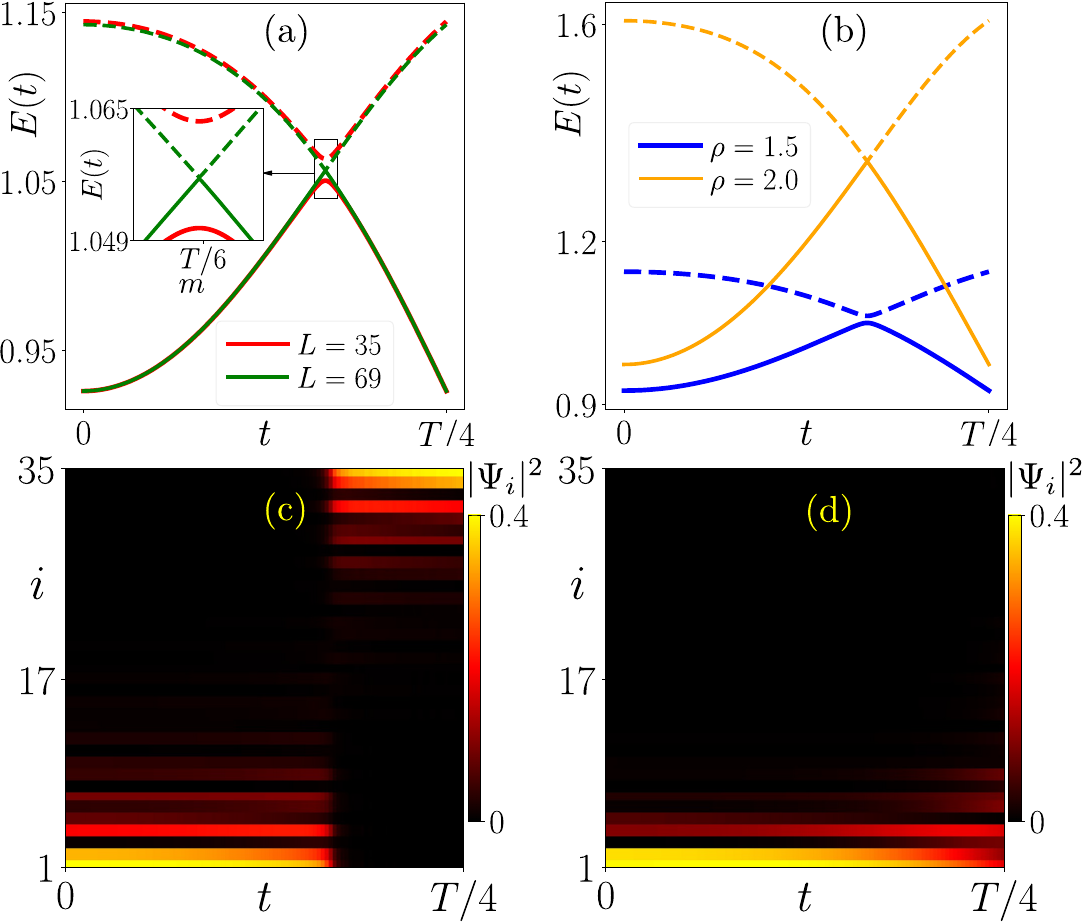}}
	\caption{(a) Eigenvalue evolution $E(t)$ of the winding state (solid) and the next excited state (dashed) as a function of time $t$ for the $C_8$ FC with $L=F_8+1=35$ (red) and $L=2F_8+1=69$ (green). Inset shows zoomed-in excitation gap. Here $\rho=1.5$. (b) Same as (a) but for different hopping ratios $\rho=1.5$ (blue) and $2.0$ (orange) for the FC $L=F_8+1=35$. (c) Evolution of the winding state as a function of time $t$ and spatial position $i$ for $\rho=1.5$ for the FC $L=F_8+1=35$. (d) Same as (c) but for $\rho=2.0$. In all panels, the transfer rate is $\Omega=10^{-4}$.}
	\label{Fig:Timedyn}
\end{figure}
%~~~~~~~~~~~~~~~~~~~~~~~~~~~~~~~~~~~~~~~~~~~~~~~~~~~~~~~~~
%~~~~~~~~~~~~~~~~~~~~~~~~~~~~~~~~~~~~~~~~~~~~~~~~~~~~~~~~~

To maintain adiabatic evolution and thus achieve complete state transfer, the winding state needs to have a finite excitation gap separating it from all other, `excited' states. Otherwise, if the excitation gap is too small or the rate $\Omega$ too high, the winding state may make a transition ending up in the excited level, which results in no clean state transfer and broken adiabaticity~\cite{LonghiLandauZenerAQT2019, ChenPRL2021}. We therefore, in Figs.~\ref{Fig:Timedyn}(a,b), investigate the evolution of the energies of the winding state (solid) and the next excited state (dashed). As seen in Fig.~\ref{Fig:Timedyn}(a), the excitation gap decreases as we increase the number of repetitions of approximants from  $L=F_8+1=35$~(red) to $L=2F_8+1=69$~(green). We can instead consider other approximants, but find also then that the excitation gap decreases with increased chain length, see Appendix~\ref{App:C6C7}. We thus attribute the existence of an excitation gap to a finite size effect, also in agreement with earlier results~\cite{ChenPRL2021}. In addition, we investigate the effect of the hopping ratio $\rho$ on the excitation gap size in Fig.~\ref{Fig:Timedyn}(b). Here, the excitation gap diminishes with increasing $\rho$. We thus need to consider an appropriate system size and hopping ratio to keep a finite excitation gap and thus adiabaticity.

We illustrate the importance of adiabaticity in Figs.~\ref{Fig:Timedyn}(c,d), where we plot the transfer of the winding state for different $\rho$. Here, a smaller~(larger) value of $\rho$ yields a larger~(smaller) excitation gap and leads to a successful~(unsuccessful) FQP operation, for a given transfer rate $\Omega$. In the former case, the state gets transferred completely to the other end of the system, following an adiabatic evolution, while in the latter case, the state remains on the same side as the initial state due to loss of adiabaticity. Furthermore, the FQP operation is also robust against both onsite and bond disorder, see Appendix~\ref{App:Disorder}.
For efficient, fast, and successful on-demand FQP operations, it is therefore essential to find optimum conditions based on controllable parameters. 

\begin{figure}
	\centering
	\subfigure{\includegraphics[width=0.49\textwidth]{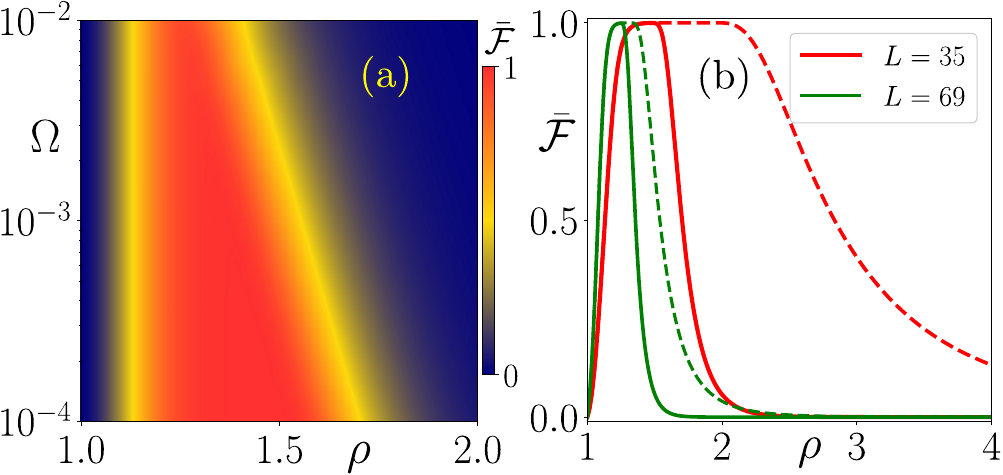}}
	\caption{(a) Weighted fidelity $\bar{\mathcal{F}}$ as a function of hopping ratio $\rho$ and rate $\Omega$ for the $C_8$ FC with $L=F_8+1=35$. The frequency axis is on a logarithmic scale. (b) $\bar{\mathcal{F}}$ as a function of $\rho$ for fixed $\Omega=10^{-4}$ for FCs with $L=F_8+1=35$~(red) and $2F_8+1=69$~(green) for the one- (solid) and two-step (dashed) transfer protocols. 
	}
	\label{Fig:PhaseDiagram}
\end{figure}

To capture the performance of the FQP quantitatively, we define a weighted fidelity $\bar{\mathcal{F}}$ (see Appendix~\ref{App:Fidelity}), which measures the overlap between the time-evolved state $\psi(t_f)$ at the end of the protocol at $t=t_f$ and a target state $\psi_{\rm T}=\psi^{\phi_2}_w$, \ie the winding state for $\phi_2$ (orange state in Fig.~\ref{Fig:FQCbackground}(c)). The fidelity is also weighted by a factor $\alpha$ that ensures that the target state is an edge state localized at the right end of the chain, see Eq.~\eqref{Eq:Fidelity} in Appendix~\ref{App:Fidelity}. Thus, $\bar{\mathcal{F}} \simeq 1$ indicates a successful transfer of states to the other end of the FC. We plot $\bar{\mathcal{F}}$ in Fig.~\ref{Fig:PhaseDiagram}(a) as a function of hopping ratio $\rho$ and rate $\Omega$ for the $C_8$ FC. For $\rho = 1$ the fidelity is zero since the system is conventional 1D metal with no edge states. But as soon as the system becomes a quasicrystal, the fidelity is increased.
We observe that $\bar{\mathcal{F}}$ is close to $1$ for a broad range of parameters, highlighting a large flexibility in the operational window of the FQP.
Especially keeping $1<\rho<1.6$ allows for very flexible transfer rates. 
This observation aligns with Fig.~\ref{Fig:Timedyn}(b), where we show the decrease of gap size for even larger hopping ratios. We note that, while the weighted fidelity $\bar{\mathcal{F}}$ signals edge-localization of the final state, it does not, however, tell how good the localization is. We therefore also compute the inverse participation ratio to capture edge-localization properties of the pumped state, see Appendix~\ref{App:IPR}. The robust and on-demand dynamic transfer of the winding state by switching only two hoppings in the whole system for a wide range of parameters is one of the main findings in this work. 

%~~~~~~~~~~~~~~~~~~~~~~~~~~~~~~~~~~~~~~~~~~~~~~~~~~~~~~~~~
%~~~~~~~~~~~~~~~~~~~~~~~~~~~~~~~~~~~~~~~~~~~~~~~~~~~~~~~~~
%~~~~~~~~~~~~~~~~~~~~~~~~~~~~~~~~~~~~~~~~~~~~~~~~~~~~~~~~~
%~~~~~~~~~~~~~~~~~~~~~~~~~~~~~~~~~~~~~~~~~~~~~~~~~~~~~~~~~

We also showcase results for a two-step protocol. In this two-step protocol, instead of changing both hoppings in a unit cell of $C_8$ simultaneously, we change them adiabatically in two separate steps, first $J_1$ and then $J_{N_B}$, see Eq.~\eqref{Eq:TwoStepHoppingChangeMainText} in Appendix~\ref{App:TransferProtocol} and also Appendix~\ref{App:TwoStepProtocol}. We compare $\bar{\mathcal{F}}$ for the one- and two-step protocols in Fig.~\ref{Fig:PhaseDiagram}(b) as a function of $\rho$. As clearly seen, $\bar{\mathcal{F}}$ remains close to $1$ over a larger range of $\rho$ values for the two-step protocol. The two-step protocol also supports more rapid change of parameters (higher rate $\Omega$) than the one-step protocol. We attribute this improvement to the increase and nature of the excitation gap separating the winding state from excited states, see Appendix~\ref{App:TwoStepProtocol}. Thus, the two-step protocol enhances the FQP flexibility by adapting the transfer process.

The pumping of winding states in the FQP demonstrates that it is possible to fully transfer states from one end of the system to the other by only minimally changing parameters by leveraging the unique quasiperiodic properties of the FC. This makes our FQP prescription highly suitable for experimental realization, as it requires significantly less tuning compared to conventionally studied (periodic)  SSH model-based pumps~\cite{LangNQI2017,LonghiPRB2019,LonghiLandauZenerAQT2019,ZhengLiNaPRA2020,QiLuPRA2020,DAngelisPRR2020,PalaiodimopoulosPRA2021,CaoPRA2021,YuanAPLPh2021,QiPRRTR2021,WangPRA2022,ZhengLiNaYiPRApp2022,WeijiePRA2022,WangDaWeiPRA2023,ZhaonppjQuantumInf2023,Zurita2023fastquantumtransfer,RomeroPRApp2024,HanJinXuanPRApp2024,TianPRB2024,Fernandez2024,WangDaWeiPRA2024,ZhengCJP2025} or previous FC protocols~\cite{KrausPumpingPRL2012,VerbinPump2015,SinghPRA2015}. Furthermore, we next demonstrate how the FQP can be used as a quantum bus between two distant qubits. Moreover, in Appendix~\ref{App:Onsitemodel}, we discuss a different type of FC, where the onsite chemical potentials instead belong to a Fibonacci sequence, while bond strengths remain constant throughout the chain, and replicate our results.

%====================================
\section{State transfer and maximal entanglement between distant qubits} \label{Sec:StateTransfer}
%====================================
%--------------------------------------------------------------
% \vspace{2mm}
% \noi \textbf{State transfer and maximal entanglement between distant qubits}
%--------------------------------------------------------------

%~~~~~~~~~~~~~~~~~~~~~~~~~~~~~~~~~~~~~~~~~~~~~~~~~~~~~~~~~
%~~~~~~~~~~~~~~~~~~~~~~~~~~~~~~~~~~~~~~~~~~~~~~~~~~~~~~~~~
\begin{figure*}
	\centering
	\subfigure{\includegraphics[width=0.98\textwidth]{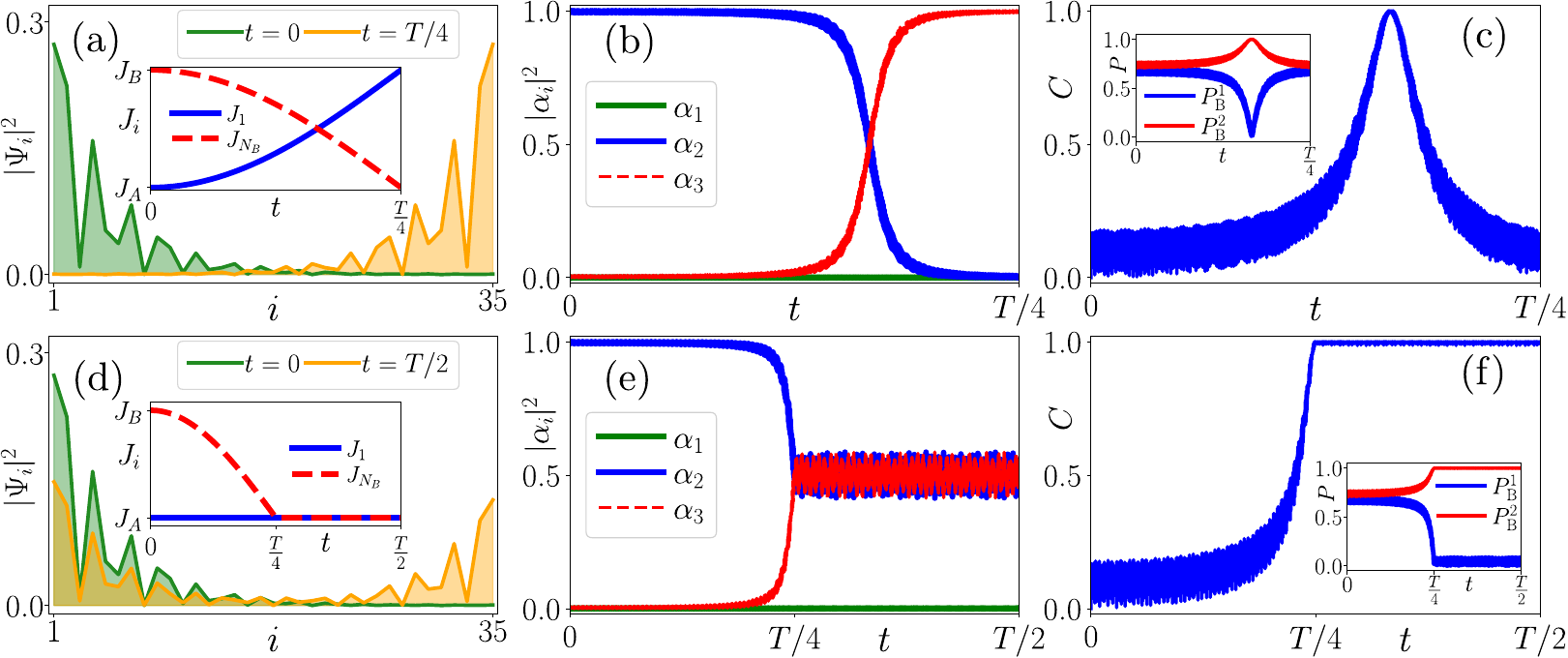}}
	\caption{(a) Spatially resolved amplitude of the winding state at initial time $t=0$~(green) and final time $t=t_f$~(orange) using the transfer protocol in inset with changes only to the hopping amplitudes $J_1$ and $J_{N_B}$. (b) Amplitudes $\lvert \alpha_{1,2,3}(t) \rvert^2$ of the time-evolved many-body state for initial state $\ket{101}$ as a function of time. (c) Concurrence $C$ as a function of time for the same initial state as in (b). Inset shows the probability of obtaining the two Bell states. (d-e) Repeats (a-c), but changes only one of the hopping amplitudes in the unit cell. Here we use the $C_8$ FC with $L=F_8+1=35$ and $\rho=1.4$, while $g=1.0$, $\Omega=10^{-4}$, and $E_l=E_r=0.0$.}
	\label{Fig:QubitCoupling}
\end{figure*}
%~~~~~~~~~~~~~~~~~~~~~~~~~~~~~~~~~~~~~~~~~~~~~~~~~~~~~~~~~
%~~~~~~~~~~~~~~~~~~~~~~~~~~~~~~~~~~~~~~~~~~~~~~~~~~~~~~~~~

\noi Next, we demonstrate the potential of the FQP for quantum information science. Specifically, we aim to employ the FQP as a quantum bus to control and mediate quantum information and entanglement between distant qubits, thereby enabling general-purpose quantum information processing. To this end, we couple the FQP to two outer qubits, $Q_1$ and $Q_2$, with a coupling strength $g$, as schematically illustrated in Fig.~\ref{Fig:Schematics}. Although the full Hamiltonian of the system in the many-body basis is considerably large, the problem can be effectively simplified by assuming that the outer qubits interact only with the winding state. Such an approximation has previously successfully been employed for matching energies between a qubit and the outer state~\cite{ChenSciRep2016}. We therefore approximate the full system with an effective three-site Hamiltonian, $\tilde{H}_{\rm MB}(t)$, see Eq.~\eqref{Eq:MBHam} in Appendix~\ref{App:EffectiveQubitHam}. Employing $\tilde{H}_{\rm MB}(t)$, and only considering the even occupation sector (see Appendix~\ref{App:EffectiveQubitHam}), we obtain the time evolved state: $\tilde{\Psi}_{\rm MB}(t)=\alpha_1(t) \ket{110}+\alpha_2(t) \ket{101}+\alpha_3(t) \ket{011}$, where we employ the basis states $\ket{n_{l} n_{r} n_{w}}$ with $n_{l}$, $n_{r}$, and $n_{w}$ representing the occupation of the left and right qubit, and the winding state, respectively, taking values $0,1$. Hence, the time-dependent weights $\alpha_{1,2,3}(t)$, carry all the quantum information stored in $\tilde{\Psi}_{\rm MB}(t)$. For instance, the concurrence between the two outer qubits, given by $C = 2 \lvert \alpha_2(t) \alpha_3(t) \rvert$, quantifies their entanglement~\cite{HillPRL1997,MazzolaPRA2010}. We further set the qubit internal energies to $E_l=E_r=0$ and the coupling strength between qubits and the FQP to $g=0.1$, but we verify that our results do not depend on this specific choice.

We employ the FQP to both transfer information between the qubits and generate entanglement. We start by showing the result of the time evolution of the winding state governed by the instantaneous FC Hamiltonian $H(t)$ (see Appendix~\ref{App:TimeEvolution}) in Figs.~\ref{Fig:QubitCoupling}(a) for the one-step transfer protocol, thus reproducing Fig.~\ref{Fig:Timedyn}(c) but now only focusing on the initial time ($t=0$, green) and final time ($t=t_f$, orange). The insets show how the outermost hoppings $J_1$ and $J_{N_B}$ are changing during the one-step transfer protocol. Figures~\ref{Fig:QubitCoupling}(b,c) then illustrate how the qubits $Q_{1,2}$ are influenced by this transfer of the winding state between the two ends of the FQP, when the system is initially prepared in the unentangled state $\ket{101}$. Thus, at $t=0$, the left~(right) qubit is occupied~(unoccupied), with the transferring state (winding state) also occupied. Specifically, in Fig.~\ref{Fig:QubitCoupling}(b) we plot the amplitudes of the weights $\lvert \alpha_{1,2,3}(t) \rvert^2$ of the time-evolved state $\tilde{\Psi}_{\rm MB}(t)$. We observe that $\lvert \alpha_{2}(t) \rvert^2$~($\lvert \alpha_{3}(t) \rvert^2$) evolve from one~(zero) at the initial time $t_0 = 0$ to zero~(one) at the final time $t_f = T/4$, while  $\lvert \alpha_{1}(t) \rvert^2 =0$. Thus, at $t_f = T/4$, the system reaches the state $\ket{011}$, indicating that the quantum information has been fully transferred from the left qubit to the right qubit. This transfer occurs as the winding state moves from the left end to the right end of the FQP, see Fig.~\ref{Fig:QubitCoupling}(a), demonstrating the capability of the FQP to operate as a quantum bus, enabling quantum information transfer between two spatially separated qubits.

The information transfer in Fig.~\ref{Fig:QubitCoupling}(b) occurs via entanglement between the two qubits, mediated by the FQP. Fig.~\ref{Fig:QubitCoupling}(c) illustrates the entanglement dynamics in terms of the time-dependence of the concurrence $C$ between the two qubits. As seen, the qubits start unentangled in $\ket{101}$, then the entanglement gradually increases as the winding state propagates across the chain. Midway through the transfer, when $J_1<J_{N_B}$ switches to $J_1>J_{N_B}$, the two qubits become maximally entangled, before again decreasing to the unentangled state $\ket{011}$, indicating completion of the quantum state transfer. At maximum concurrence, the time-evolved state also becomes identical to a Bell state. This is seen in the inset of Fig.~\ref{Fig:QubitCoupling}(c), where we plot the overlap with the two Bell states, $P_{\rm B}^{1,2} = \lvert \braket{\tilde{\Psi}_{\rm MB}(t) | \Psi_{\rm B}^{1,2}} \rvert$, with the Bell states defined as $\Psi_{\rm B}^{1,2} = \left( \ket{101} \pm \ket{011} \right)/\sqrt{2}$. In Appendix~\ref{App:Entanglement}, we provide complementary results for state transfer and entanglement dynamics for different initial states.

The limited time-window for maximum entanglement in Fig.~\ref{Fig:QubitCoupling}(c) can be improved if we stop the protocol of the FQP at the peak point of the concurrence. However, this switching point is sharp and may be difficult to control, see Appendix~\ref{App:Entanglement}. Instead, we engineer maximum entanglement by using a partial-state transfer protocol where only $J_{N_B}$ changes from $J_B \rightarrow J_A$, while $J_1$ remains fixed at $J_A$, see inset of Fig.~\ref{Fig:QubitCoupling}(d) and Eq.~\eqref{eq:HoppingChangeEntanglement} in Appendix~\ref{App:TransferProtocol} for the explicit form of this protocol. For this protocol, we set the final time to $t_f=T/2$ to show entanglement stability, but the switching of $J_{N_B}$ is complete already at $t=T/4$. In Fig.~\ref{Fig:QubitCoupling}(d), we show that the initial state (green) does not get completely transferred from the left to the right end of the FC; rather, the final state (orange) is now localized at both ends of the chain in a superposition. As a consequence, the qubits become coupled via the FQP. We verify this in Fig.~\ref{Fig:QubitCoupling}(e), again starting with the initial state $ \ket{101}$ but now ending up with an equal superposition of the $\ket{101}$ and $ \ket{011}$ from $t=T/4$ and onward. The result is maximal concurrence $C=1$ and the Bell state $\Psi_{\rm B}^2$, as seen in Fig.~\ref{Fig:QubitCoupling}(f). This demonstrates that the FQP can be used both to transfer states between qubits and to prepare maximally entangled states between them.

%====================================
\section{Summary and Outlook} \label{Sec:Summary}
%====================================

%============================================
% \vspace{5mm}
% \noi \textbf{\Large Discussion}
% \vspace{1mm}
%============================================

\noi In this work, we show that the intriguing quasiperiodic properties of the Fibonacci chain (FC) can be successfully utilized to construct a Fibonacci quantum pump (FQP) that requires modulating only two single couplings to allow for complete state transfer of a quasicrystal winding state, likely making it the minimal possible topological pump. In particular, we demonstrate complete quantum pumping of winding states from one end of the chain to the other by solving for the full time dependence and keeping adiabatic evolution. The parameter range for successful pumping with maximal fidelity is large and highly tunable by both configuring the FC chain and the transfer protocol. Importantly, we also demonstrate that the FQP acts as a quantum bus for mediating quantum information exchange between two distant qubits, by being able to both perform perfect quantum state transfer between the qubits and generate a maximally entangled Bell state.

While we have demonstrated a FQP with quantum state transfer and maximal entanglement generation, we envision that more general quasicrystals may possibly have even more improved properties, for example, larger gaps protecting adiabatic evolution from excited states or stronger winding state localization. Moreover, optimization of the FQP operation also involves optimization of the transfer protocol. For example, achieving even faster transfer rates might be possible by specially engineered protocols, such as following the framework of shortcuts to adiabaticity~\cite{GreentreePRB2004,GueryOdelinRMP2019}. Still, our one- and two-step protocols are highly beneficial and may actually be easier to implement, as we only change two couplings in a straightforward simultaneous or successive manner, resulting in very simple protocols.

Our work provides the necessary proof-of-principle to investigate the experimental realization in quasiperiodic systems, but also paves the way for many other interesting further studies. For instance, the impact of the environment on the FQP warrants further investigation. Coupling to the environment can lead to dissipation, which may induce decoherence during the transfer protocol and modify the FQP operation regime~\cite{FedorovaNC2020,DreonNature2022}. In contrast, dissipation has also been exploited to induce topologically protected states~\cite{Diehl2008,Diehl2011,BudichPRA2015,GhoshDissipation2024} and have even been shown to improve the state transfer in some cases~\cite{Verstraete2009,TangPRR2025}.

Our FQP setup is likely possible to implement in multiple different experimental setups. One of the potential platforms is a lattice of superconducting resonators~\cite{SplitthoffPRR2024}, where modifying couplings between different resonators using an electrostatic gate potential has already been employed to engineer the SSH model. Instead, utilizing the FQP requires only one or two gate potential changes, while for the SSH model, the number of changes scales with the total length of the chain.
Another potential setup may be fermionic simulators made out of arrays of coherently coupled Josephson junctions~\cite{johannsen2024fermionic}. Also, photonic waveguide arrays can be employed to realize the FQP, where a hopping Hamiltonian can be engineered, with longer~(shorter) separation between two waveguides effectively describing weaker~(stronger) bond strengths. In this latter setup, the propagation direction of light mimics the time axis. However, unlike previous implementations~\cite{KrausPumpingPRL2012,VerbinPump2015}, requiring changing all the inter-waveguide distances along the propagation direction to pump a state, our approach simplifies the setup substantially by only requiring adjustment of the outermost waveguides, keeping all the others fixed. Thus, our work highlights the advantages of quasiperiodic systems for long-range entanglement and quantum state transfer, potentially facilitating quantum experiments and simplifying quantum algorithms.

%=====================================================
\subsection*{Data Availability}
%======================================================
The data that support the findings of this article are openly available at Ref.~\cite{Ghoshdata}.

\subsection*{Acknowledgments}

\noi We acknowledge Aksel Kobiałka and Vimalesh K.~Vimal for the helpful discussions. We acknowledge financial support from the Swedish Research Council (Vetenskapsrådet) Grant No.~2022-03963 and the European Research Council (ERC) under the European Union’s Horizon 2020 research and innovation programme (ERC-2022-CoG, Grant agreement No.~101087096). Part of the computations were enabled by resources provided by the National Academic Infrastructure for Supercomputing in Sweden (NAISS), partially funded by the Swedish Research Council through grant agreement No.~2022-06725. R.S.S. acknowledges funding from the Spanish Comunidad de Madrid (CM) ``Talento Program'' (Project No. 2022-T1/IND- 24070), and the Spanish Ministry of Science, innovation, and Universities through Grant No. PID2022-140552NA-I00.

\appendix

%======================================================
% \vspace{5mm}
% \noi \textbf{\Large Methods}
% \vspace{1mm}
%======================================================

%---------------------------------
\section{Methods} \label{App:Method}
%---------------------------------

%---------------------------------
\subsection{Transfer protocols} \label{App:TransferProtocol}
%---------------------------------

\noi To transfer a state between the two ends of the FC, we dynamically change the hopping parameters. In Figs.~\ref{Fig:Timedyn}, \ref{Fig:PhaseDiagram}, and \ref{Fig:QubitCoupling}(a-c), we employ the one-step protocol. Explicitly, the one-step protocol reads
\begin{align}
    V(t) \!= \! \left\{ \!J_i^{\phi_1} \!\right\} \!+\! \left(\! \left\{ \!J_i^{\phi_2} \!\right\}\!-\! \left\{ \! J_i^{\phi_1} \! \right\} \! \right) \! \left(1\!-\!\cos \Omega t \right), \  t \in [0,T/4] ,
    \label{Eq:driveprotcol}
\end{align}
where $\left\{ J_i^{\phi_{1,2}} \right\}$ represents the hopping sequence corresponding to phason angles $\phi_{1,2}$, while $\Omega$ is the transfer rate. Further, $T=2\pi/\Omega$ and we set $t_f=T/4$ to the time where the process is completed. Note that the transfer protocol $V(t)$ only changes the two outermost hoppings in the FC (or two hoppings per approximant repetition), see Fig.~\ref{Fig:FQCbackground}(b), while all other hoppings remain unchanged. Explicitly, the two outermost hoppings change as
\begin{align} 
\label{eq:J1JNB}
    J_1 (t)=&J_B -(J_B-J_A)  \cos \Omega t \ , \quad t \in [0,T/4] , \non \\
    J_{N_B} (t)=&J_A +(J_B-J_A)  \cos \Omega t \ , \quad t \in [0,T/4],
\end{align}
illustrated in the inset of Fig.~\ref{Fig:QubitCoupling}(a). In FCs repeating a single approximant $l$ times, e.g., in Figs.~\ref{Fig:Timedyn}(a) and \ref{Fig:PhaseDiagram}(b), we repeat Eq.~\eqref{eq:J1JNB} in each approximant, resulting in $2l$ hoppings being changed. We illustrate such a change of multiple hoppings for a repeated chain in Fig.~\ref{Fig:C6C7}.

The two-step protocol we employ in Fig.~\ref{Fig:PhaseDiagram}(b) instead explicitly reads as
\begin{align}
    J_1 (t)=&J_B -(J_B-J_A)  \cos \Omega t \ , \quad t \in [0,T/4] , \non \\
    J_{N_B} (t)=&J_B +(J_B-J_A)  \cos \Omega t \ , \quad t \in [T/4,T/2].
    \label{Eq:TwoStepHoppingChangeMainText}
\end{align}
For this two-step protocol, we instead set $t_f=T/2$ to the time where the process is completed. The hopping changes in the two-step protocol are shown in Fig.~\ref{Fig:TwoStep} in Appendix~\ref{App:TwoStepProtocol}.

Finally, to generate entanglement in Figs.~\ref{Fig:QubitCoupling}(d,e), we use the partial-transfer protocol where we only change one of the hoppings as
\begin{align}
    J_{N_B} (t)=&J_A +(J_B-J_A)  \cos \Omega t \ , \quad t \in [0,T/4],
    \label{eq:HoppingChangeEntanglement}
\end{align}
while $J_1$ remains fixed at $J_A$. For this protocol, the hopping change is completed at $t=T/4$, but in Fig.~\ref{Fig:QubitCoupling}(d-e), we still study the system until $t=T/2$ to show that we retain maximal entanglement.
%

%---------------------------------
\subsection{Time evolution} \label{App:TimeEvolution}
%---------------------------------
% \vspace{2mm}
% \noi \textbf{Time evolution}

\noi For the one-step transfer protocol, we obtain the instantaneous Hamiltonian $H(t)$ at time $t$ by using the hopping sequence $V(t)$ in Eq.~\eqref{Eq:driveprotcol}, such that $H(t)=H \left[ V(t) \right]$, with $H(0)=H \left[ \left\{ J_i^{\phi_1} \right\} \right]$. 
To obtain time-evolved states, we construct the time-evolution operator in a time-ordered~($\mathcal{T}$) manner as $U(t,0)=\mathcal{T} \exp \left[ -i \int^t_0 H(t') dt' \right]$. We then compute the time-evolved state $\psi(t)$ as $\psi(t)=U(t,0) \psi_w(0)$, where $\psi_w(0)$ represents the winding state of the Hamiltonian $H(0)$ at initial time $t=0$, corresponding to the hopping sequence for the phason angle $\phi_1$. Then at final time $t_f = T/4$, the Hamiltonian $H(t_f)$ represents a FC with a hopping sequence generated for the phason angle $\phi_2$, where $T=2\pi/\Omega$ with transfer rate $\Omega$. Similarly, for the two-step and partial-transfer protocols, we use Eqs.~\eqref{Eq:TwoStepHoppingChangeMainText} and \eqref{eq:HoppingChangeEntanglement}, respectively, to obtain $H(t)$ and thereby their time-evolution operators $U(t,0)$.

%---------------------------------
\subsection{Fidelity} \label{App:Fidelity}
%---------------------------------
% \vspace{2mm}
% \noi \textbf{Fidelity}

\noi We define a weighted fidelity $\bar{\mathcal{F}}$ employing the target state $\psi_{\rm T}$ as
\begin{align}
    \bar{\mathcal{F}}= \alpha \lvert \langle \psi(t=t_f) | \psi_{\rm T} \rangle   \rvert^2 \ , 
    \label{Eq:Fidelity}
\end{align}
where $\psi_{\rm T}$ is the winding state of the FC Hamiltonian for $\phi_2$: $\psi_{\rm T}=\psi_w^{\phi_2}$, see orange state in Fig.~\ref{Fig:FQCbackground}(c). Here, $\psi(t=t_f)$ is the time-evolved winding state, while the weight $\alpha$ is computed as $\alpha= \left (\lvert \psi_{L,\rm T}\rvert^2 -\lvert\psi_{1,\rm T} \rvert^2 \right)/{\rm maximum}\left(\left\{\lvert \psi_{i,\rm T}\rvert^2 \right\}\right)$; with $\psi_{i,\rm T}$ being the elements of $\psi_{\rm T}$ at lattice sites $i$, with $L$ the last site of the chain. Including  $\alpha$ into a weighted fidelity ensures that the target state is an edge state with a maximum weight at the right end of the chain. We find that this is important close to $\rho \simeq 1$, where the chain starts to resemble a 1D metal with an almost delocalized winding state.

%---------------------------------
\subsection{Effective qubit-FQP-qubit setup} \label{App:EffectiveQubitHam}
%---------------------------------
% \vspace{2mm}
% \noi \textbf{Effective qubit-FQP-qubit setup}

\noi We model the qubit-FQP-qubit system as an effective three-site system, with the middle site being constructed out of the time-evolved winding state $\psi(t)$ that we numerically obtain by evolving $\psi_w(0)$ using $U(t,0)$. When the energies of the qubits are close to that of the winding state $\psi(t)$, the qubits interact only with the winding state, and we can then safely work with this effective three-site Hamiltonian~\cite{ChenSciRep2016}. In practice, this means we need to numerically extract the energy of the winding state and its weight on the first site, $i=1$, and last site, $i=L$, of the FC at each time $t$. These weights then determine the coupling with the corresponding qubit. In this description, the effective Hamiltonian of the system reads~\cite{ChenSciRep2016}
\begin{align}
    \tilde{H}=& E_w (t) c_w^\dagger c_w + E_{l} d_l^\dagger d_l  + E_{r} d_r^\dagger d_r \non \\
    &+ g a_l(t) d_l^\dagger c_w + g a_r(t) c_w^\dagger d_r + {\rm H.c.} , 
    \label{Eq:HamqubitFCcoupling}
\end{align}
where $E_w(t)~(c_w^\dagger)$, $E_l~(d_l^\dagger)$, and $E_w~(d_r^\dagger)$ are the energies (electronic creation operators) of the time-evolved winding state, left, and right qubits, respectively. The energy $E_w(t)$ here becomes a time-dependent parameter that we obtain from the eigenvalues of $H(t)$, see e.g.~solid curves in Figs.~\ref{Fig:Timedyn}(a,b). The qubit energies $E_{l,r}$ should, in principle, be chosen close to $E_w(t)$. However, once we make the approximation that the qubits are only interacting with the time-evolved winding state and construct the Hamiltonian in Eq.~\eqref{Eq:HamqubitFCcoupling}, we find that the choices for $E_{l,r}$ do not matter as much and we therefore set them to zero for simplicity. We verify that different values of $E_{l,r}$ do not alter our discussion. Moreover, $a_l(t)=\lvert \psi_{i=1}(t) \rvert^2~\left(a_r(t)=\lvert \psi_{i=L}(t) \rvert^2 \right)$ is the weight of the time-evolved winding state at the left~(right) end of the chain, which we extract numerically at each time, while $g$ is a free parameter controlling the coupling strength between the winding state and the qubit.

To solve the Hamiltonian Eq.~\eqref{Eq:HamqubitFCcoupling}, we consider a basis $\ket{n_{l} n_{r} n_{w}}$, where $n_{l}$, $n_{r}$, and $n_{w}$ represent the occupation of the left qubit state, right qubit state, and the time-evolved winding state, respectively, taking values $n=0,1$. Since the Hamiltonian in Eq.~\eqref{Eq:HamqubitFCcoupling} preserves the number of particles, it does not couple even and odd number of particle sectors, and we can focus on only the even number of particles sector, while an analysis of odd number of particles sector gives the same results. We thus use the even sector basis: $\left(\left\{ \ket{101}, \ket{011}, \ket{110} \right\}^{\tilde{T}} \right)$; with $\tilde{T}$ representing transpose, with the many-body Hamiltonian in this basis reading as
\begin{align}
    \tilde{H}_{\rm MB}(t)=
    \begin{pmatrix}
         E_{l}+E_{r} &  g a_r(t) &   g a_l(t) \\
        g a_r(t) &   E_{l} + E_w(t) & 0 \\
        g a_l(t) & 0 &  E_{r}+E_w(t)
    \end{pmatrix} \ .
    \label{Eq:MBHam}
\end{align}
We here omit the $\ket{000}$ state as there is no term in the Hamiltonian associated with this state. We employ the Hamiltonian in Eq.~\eqref{Eq:MBHam} to construct the time-evolution operator $\tilde{U}_{\rm MB}(t,0)=\mathcal{T} \exp \left[ -i \int^t_0 \tilde{H}_{\rm MB}(t') dt' \right]$ and compute the time-evolved many-body state as $\tilde{\Psi}_{\rm MB}(t)=\tilde{U}_{\rm MB}(t,0)\tilde{\Psi}_{\rm MB}(0)$. This time-evolved state can always  be expressed as $\tilde{\Psi}_{\rm MB}(t) = \alpha_1(t)\ket{110} + \alpha_2(t)\ket{101} + \alpha_3(t)\ket{011}$. Thus, the time-dependent weights $\alpha_{1,2,3}(t)$ govern all properties, including the dynamics of the quantum state transfer and the qubit-qubit entanglement distribution. 

%======================================================
\section{Fibonacci quantum pump using different Fibonacci chains} \label{App:C6C7}
%======================================================

%~~~~~~~~~~~~~~~~~~~~~~~~~~~~~~~~~~~~~~~~~~~~~~~~~~~~~~~~~
%~~~~~~~~~~~~~~~~~~~~~~~~~~~~~~~~~~~~~~~~~~~~~~~~~~~~~~~~~
\begin{figure}
	\centering
	\subfigure{\includegraphics[width=0.48\textwidth]{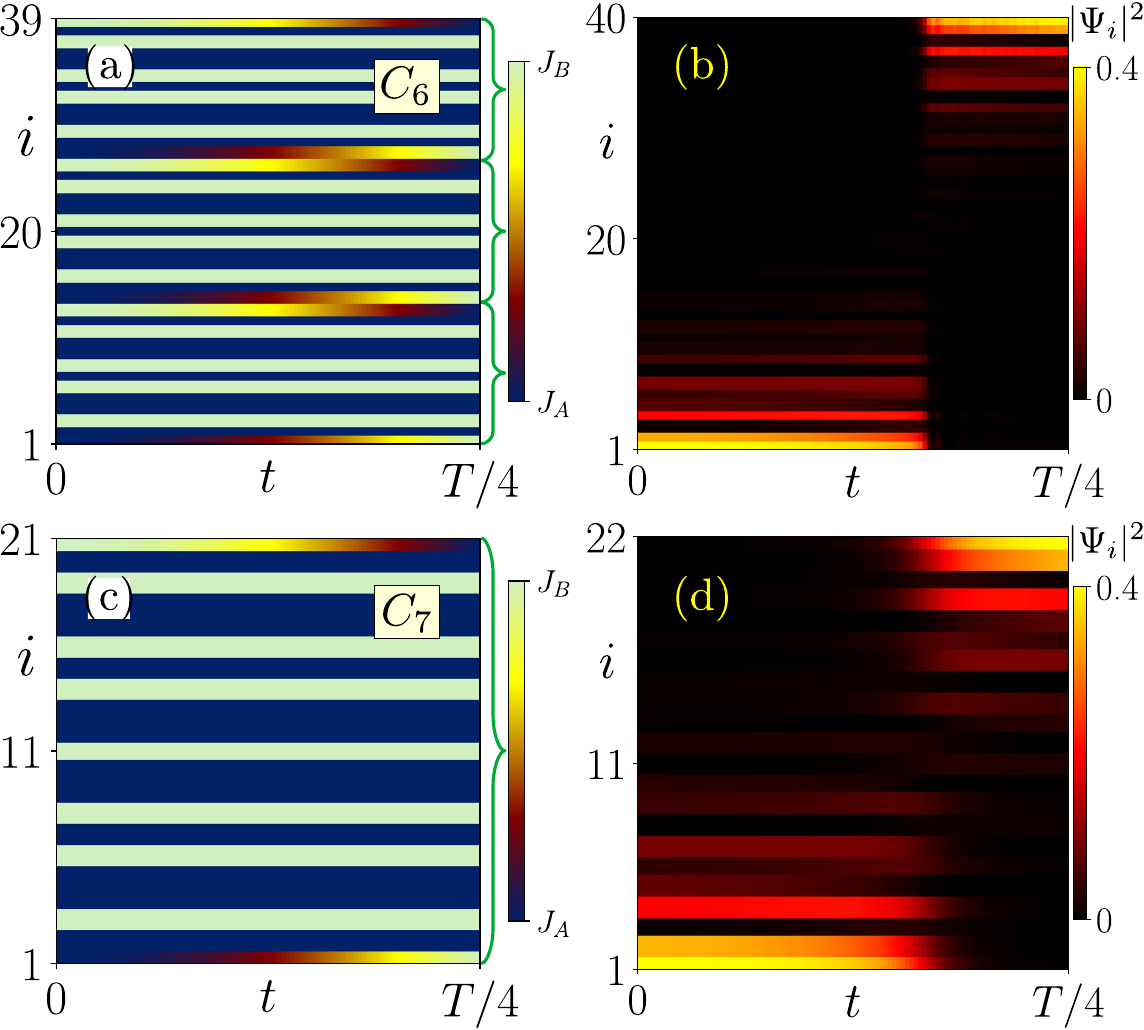}}
	\caption{(a) Dynamical change of hoppings as a function of time $t$ for the $C_6$ FC with $l=3$ repetitions such that $L=3F_6+1=40$, with colors representing the amplitude of the hopping, with $J_A$ (blue) and $J_B$ (green). Curly brackets represent one unit cell of the FC.  
    (b) Pumping of the winding state as a function of lattice site $i$ and time $t$ for the dynamical change of hoppings in (a). (c-d) Repeats (a-b) but for the $C_7$ FC with $l=1$ such that $L=F_7=22$. Here $\rho=1.5$, $\Omega=10^{-4}$.
	}
	\label{Fig:C6C7}
\end{figure}
%~~~~~~~~~~~~~~~~~~~~~~~~~~~~~~~~~~~~~~~~~~~~~~~~~~~~~~~~~
%~~~~~~~~~~~~~~~~~~~~~~~~~~~~~~~~~~~~~~~~~~~~~~~~~~~~~~~~~

In the main text, we discuss the Fibonacci quantum pump~(FQP) employing the $C_8$ Fibonacci chain~(FC), mainly for one repetition ($l=1$), but also for two ($l=2$). Here, we demonstrate that the FQP can also easily be implemented using other approximants. To this end, we first consider the $C_6$ and $C_7$ FC approximants, but we obtain analogous results also for other approximants. We then also extract the excitation gaps for the winding states for different lengths and generations of the FCs.

The unit cell of $C_6$ FC consists of $13$ bonds ($N_B =13$). In Fig.~\ref{Fig:C6C7}(a,b) we demonstrate the same transfer protocol as in Eq.~\eqref{Eq:driveprotcol} but employing a chain comprising of three repetitions ($l = 3$) of the $C_6$ FC, such that the number of sites is $L=3 F_6+1=40$. This means that we change the two outer bonds in each unit cell, i.e.~$J_1$ and $J_{N_B}$, of the $C_6$ FC dynamically, resulting in a total of six bonds being changed. We explicitly showcase this dynamic change of the bonds in Fig.~\ref{Fig:C6C7}(a), where the unit cell is indicated by curly brackets. In Fig.~\ref{Fig:C6C7}(b), we then show how this results in a complete pumping of the winding state from one end to the other of the FC.
In Fig.~\ref{Fig:C6C7}(c,d), we similarly demonstrate the FQP for a $C_7$ FC with only one repetition such that $L=21+1=22$. We again dynamically change only the two outermost hoppings, with the change in hoppings as a function of time $t$ shown in Fig.~\ref{Fig:C6C7}(c), while Fig.~\ref{Fig:C6C7}(d) shows the pumping of the winding state. Note that, in this case, we only need to change two of the hoppings.

%~~~~~~~~~~~~~~~~~~~~~~~~~~~~~~~~~~~~~~~~~~~~~~~~~~~~~~~~~
%~~~~~~~~~~~~~~~~~~~~~~~~~~~~~~~~~~~~~~~~~~~~~~~~~~~~~~~~~
\begin{figure}
	\centering
	\subfigure{\includegraphics[width=0.48\textwidth]{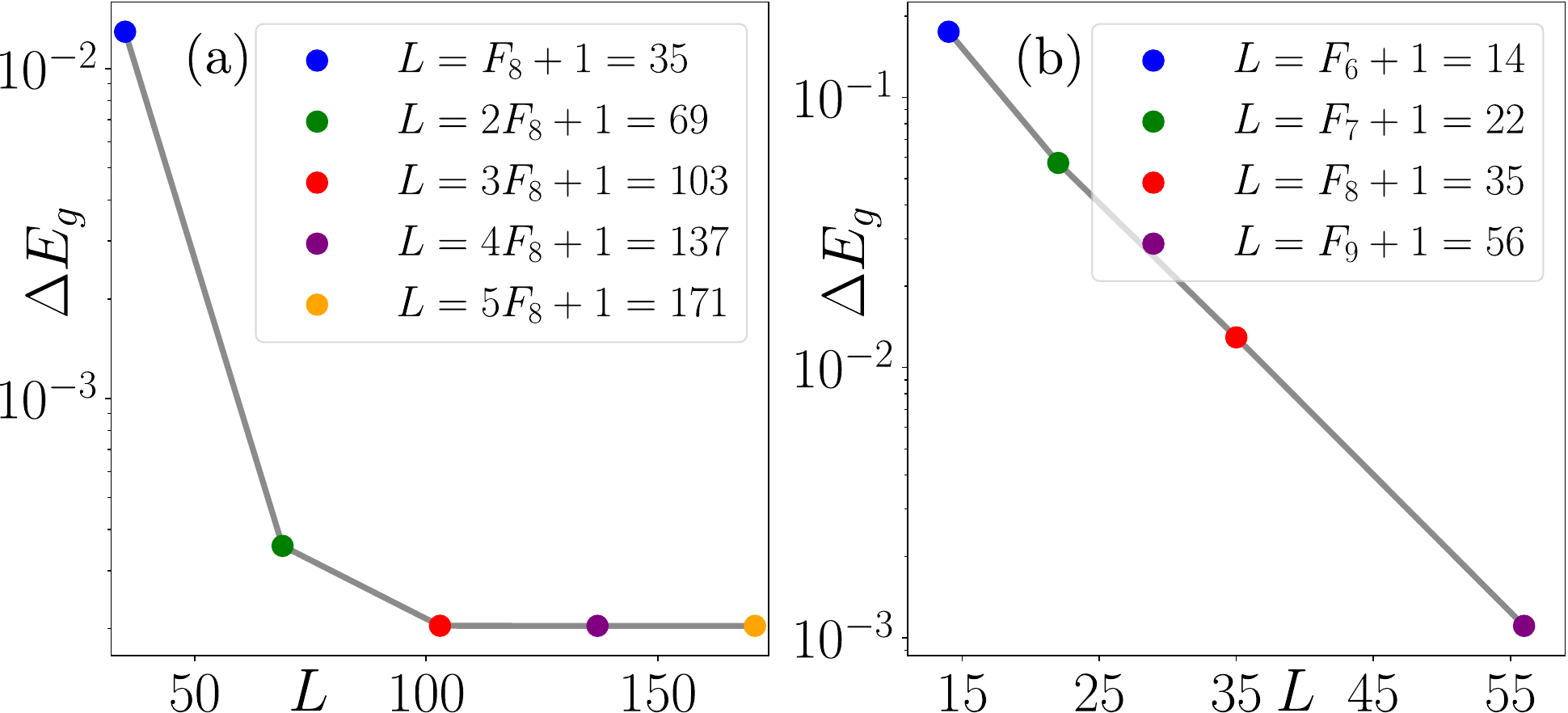}}
	\caption{(a) Excitation gap $\Delta E_g$ as a function of system sizes $L$ for multiple repetitions $l$ of the $C_8$ FC. (b) Excitation gap $\Delta E_g$ gap for a single unit cell $l=1$ of different FC approximants. Here we use $\mu=0$ and $\rho=1.5$.}
	\label{Fig:ExcitationGap}
\end{figure}
%~~~~~~~~~~~~~~~~~~~~~~~~~~~~~~~~~~~~~~~~~~~~~~~~~~~~~~~~~
%~~~~~~~~~~~~~~~~~~~~~~~~~~~~~~~~~~~~~~~~~~~~~~~~~~~~~~~~~

The results in the main text and here in Fig.~\ref{Fig:C6C7} illustrate some trade-offs for the choice of FC for the FQP. 
Using a higher generation of the FC can be beneficial, as then we can reach a longer chain, or transfer distance, with fewer hoppings that need to be changed. For instance, a single unit cell ($l= 1$) only requires changing the two outermost hoppings, making it likely a minimal topological quantum pump. In contrast, shorter approximants of the FC require us to make two changes per unit cell ($l$).
Still, repeating unit cells might be desirable in order to create a more localized winding state. This is seen in Fig.~\ref{Fig:C6C7}, where the winding state in (d) is notably less localized than in (b). 

On the other hand, while longer FCs have a more strongly localized winding state, the excitation gap, $\Delta E_g$, that separates the winding state from the next excited state during time evolution decreases with the increase in system size. We demonstrate this feature in Fig.~\ref{Fig:ExcitationGap}. In particular, Fig.~\ref{Fig:ExcitationGap}(a) shows the excitation gap $\Delta E_g$ for different system sizes $L$ for the FC approximant $C_8$. We observe that the excitation gap notably decreases with system size. This figure complements Fig.~\ref{Fig:Timedyn}(a) in the main text. In Fig.~\ref{Fig:ExcitationGap}(b), we additionally illustrate the excitation gap $\Delta E_g$ for a single unit cell ($l=1$) of different Fibonacci approximants. For longer approximants, we again have a longer chain, and the excitation gap decreases. Thus, for longer FC chains, we need to have a smaller rate that changes the bonds more slowly to be able to adiabatically pump the winding state. Depending on the demands for speed, localization, fidelity, and simplicity of implementation, different FCs might thus best fulfill the desired requirements.

%======================================================
\section{Robustness against disorder} \label{App:Disorder}
%======================================================
In the main text, we discuss the phase diagram for successful state transfer in Fig.~\ref{Fig:PhaseDiagram} in terms of the weighted fidelity $\bar{\mathcal{F}}$. However, we there do not consider any disorder. Here, we study the effect of disorder on the operation of the FQP.

We again focus on the $C_8$ FC, as in the main text, but now also add disorder in either the onsite $(\mu+W_i)$ or hopping $(J_i+W_i)$ terms in the Hamiltonian in Eq.~\eqref{Eq:HamHopping} in the main text. Here, $W_i$ represents disorder, which we assume to be uniformly distributed using $W_i \in \left[-u/2,u/2 \right]$, with $u$ representing the disorder strength in units of $J_A$. We focus on the weighted fidelity $\bar{\mathcal{F}}$, an observable that quantifies successful FQP operation. We compute the weighted fidelity $\bar{\mathcal{F}}$ for different random disorder configurations and then take an average over $50$ disorder configurations for each $u$.

\begin{figure}
	\centering
	\subfigure{\includegraphics[width=0.48\textwidth]{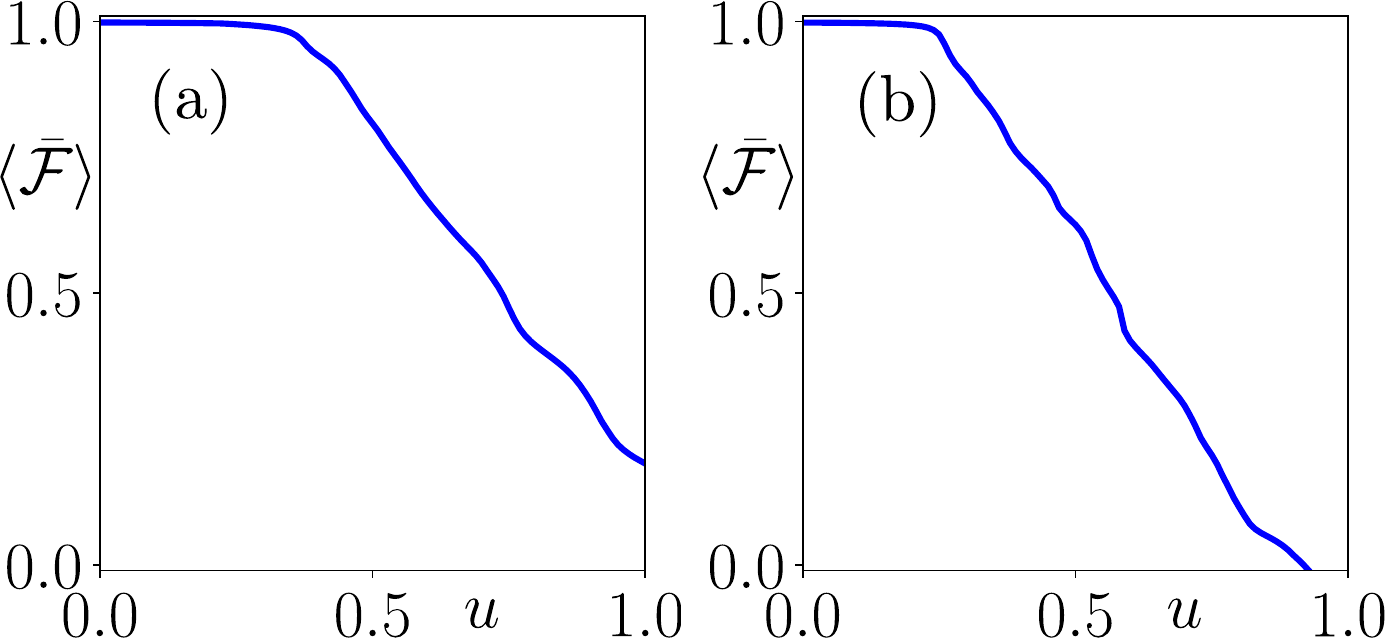}}
	\caption{Disorder-averaged weighted fidelity $\langle \bar{\mathcal{F}} \rangle$ as a function of the disorder strength $u$ for (a) onsite and (b) hopping disorder. Here we use the $C_8$ FC with $L=F_8+1=35$, $\mu=0$, $\rho=1.4$, and $\Omega=10^{-4}$ and average over $50$ random disorder configurations for each $u$.
	}
	\label{Fig:DisorderFidelity}
\end{figure}

In Fig.~\ref{Fig:DisorderFidelity}, we plot the disorder-averaged weighted fidelity $\langle \bar{\mathcal{F}} \rangle$ as a function of the disorder strength $u$ for onsite (a) and hopping (b) disorder, respectively. We find that the weighted fidelity $\langle \bar{\mathcal{F}} \rangle$ remains close to $1$ for small to moderate disorder strengths, but eventually diminishes as we increase the disorder strength. For hopping disorder, the fidelity decreases somewhat faster, which we attribute to it breaking quasiperiodicity. Nevertheless, these results show that the FQP displays remarkable robustness against disorder.

%~~~~~~~~~~~~~~~~~~~~~~~~~~~~~~~~~~~~~~~~~~~~~~~~~~~~~~~~~
%~~~~~~~~~~~~~~~~~~~~~~~~~~~~~~~~~~~~~~~~~~~~~~~~~~~~~~~~~
%~~~~~~~~~~~~~~~~~~~~~~~~~~~~~~~~~~~~~~~~~~~~~~~~~~~~~~~~~
%~~~~~~~~~~~~~~~~~~~~~~~~~~~~~~~~~~~~~~~~~~~~~~~~~~~~~~~~~

%======================================================
\section{Localization properties} \label{App:IPR}
%======================================================

%~~~~~~~~~~~~~~~~~~~~~~~~~~~~~~~~~~~~~~~~~~~~~~~~~~~~~~~~~
%~~~~~~~~~~~~~~~~~~~~~~~~~~~~~~~~~~~~~~~~~~~~~~~~~~~~~~~~~
\begin{figure}
	\centering
	\subfigure{\includegraphics[width=0.48\textwidth]{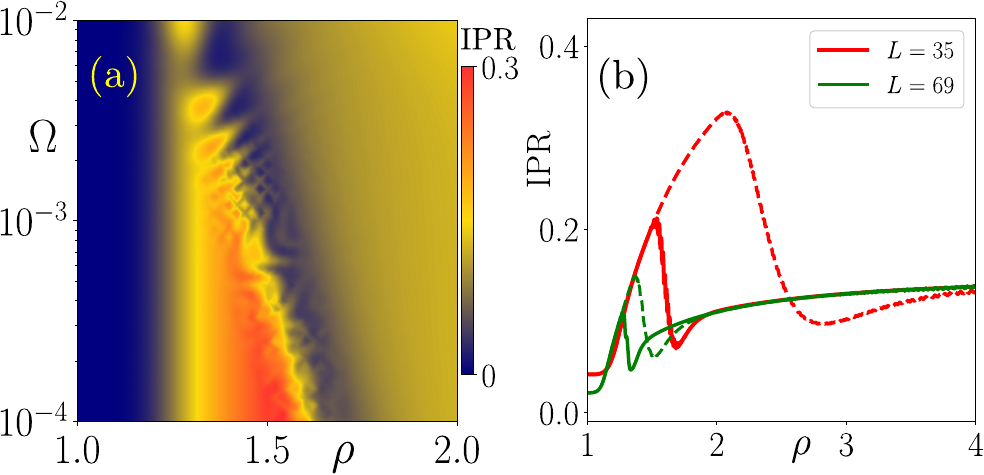}}
	\caption{(a) IPR as a function of hopping ratio $\rho$ and rate $\Omega$ for $C_8$ FC with $L=F_8+1=35$ using the one-step protocol. The frequency axis is on a logarithmic scale. (b) IPR as a function of $\rho$ for a fixed $\Omega=10^{-4}$ for FCs with $L=F_8+1=35$~(red) and $L=2F_8+1=69$~(green) for the one- (solid) and two-step (dashed) transfer protocols.
	Same systems and plots as Fig.~\ref{Fig:PhaseDiagram} in the main text, but for the IPR instead of the weighted fidelity.}
	\label{Fig:IPR}
\end{figure}
%~~~~~~~~~~~~~~~~~~~~~~~~~~~~~~~~~~~~~~~~~~~~~~~~~~~~~~~~~
%~~~~~~~~~~~~~~~~~~~~~~~~~~~~~~~~~~~~~~~~~~~~~~~~~~~~~~~~~
In Fig.~\ref{Fig:PhaseDiagram}(a), we study the phase diagram for the one-step protocol in terms of the weighted fidelity. By choosing to study a weighted fidelity, expressed in Eq.~\eqref{Eq:Fidelity}, we know that the final state is mainly located at the right end of the chain, but it does not give us any information about the degree of localization. Here, we provide additional information about the localization properties by studying the inverse participation ratio~(IPR).

The IPR for the time-evolved winding state is defined as ${\rm IPR}=\lvert \psi(t=t_f) \rvert^4$. A (lower)~higher value of IPR thus indicates a (de)localized state. We plot the IPR as a function of the hopping ratio $\rho$ and the rate $\Omega$ in Fig.~\ref{Fig:IPR}(a), i.e.~analogous to the phase diagram for the weighted fidelity in Fig.~\ref{Fig:PhaseDiagram}(a) in the main text.  Our results illustrate that if a sharply localized state is desired, we may resort to a higher value of hopping ratio $\rho$, but in that case, the parameter change rate should be a bit slower. In Fig.~\ref{Fig:IPR}(b) we also repeat Fig.~\ref{Fig:PhaseDiagram}(b) in the main text but for the IPR, by comparing the IPR for one-step (solid) and two-step (dashed) protocols for different system sizes $L=F_8+1=35$~(red) and $2F_8+1=69$~(green) as a function of the hopping ratio $\rho$. As we increase the chain length, we find a smaller range with higher IPR, while for the two-step protocol, the range for a higher IPR is larger compared to the one-step protocol, both resulting in overall agreement with the behavior of the weighted fidelity. Thus, these results show that the IPR and the weighted fidelity largely track each other and that it is possible to design the FC and the transfer protocol to reach a high weighted fidelity and IPR.

%======================================================
\section{Two-step protocol} \label{App:TwoStepProtocol}
%======================================================

%~~~~~~~~~~~~~~~~~~~~~~~~~~~~~~~~~~~~~~~~~~~~~~~~~~~~~~~~~
%~~~~~~~~~~~~~~~~~~~~~~~~~~~~~~~~~~~~~~~~~~~~~~~~~~~~~~~~~
\begin{figure}
	\centering
	\subfigure{\includegraphics[width=0.48\textwidth]{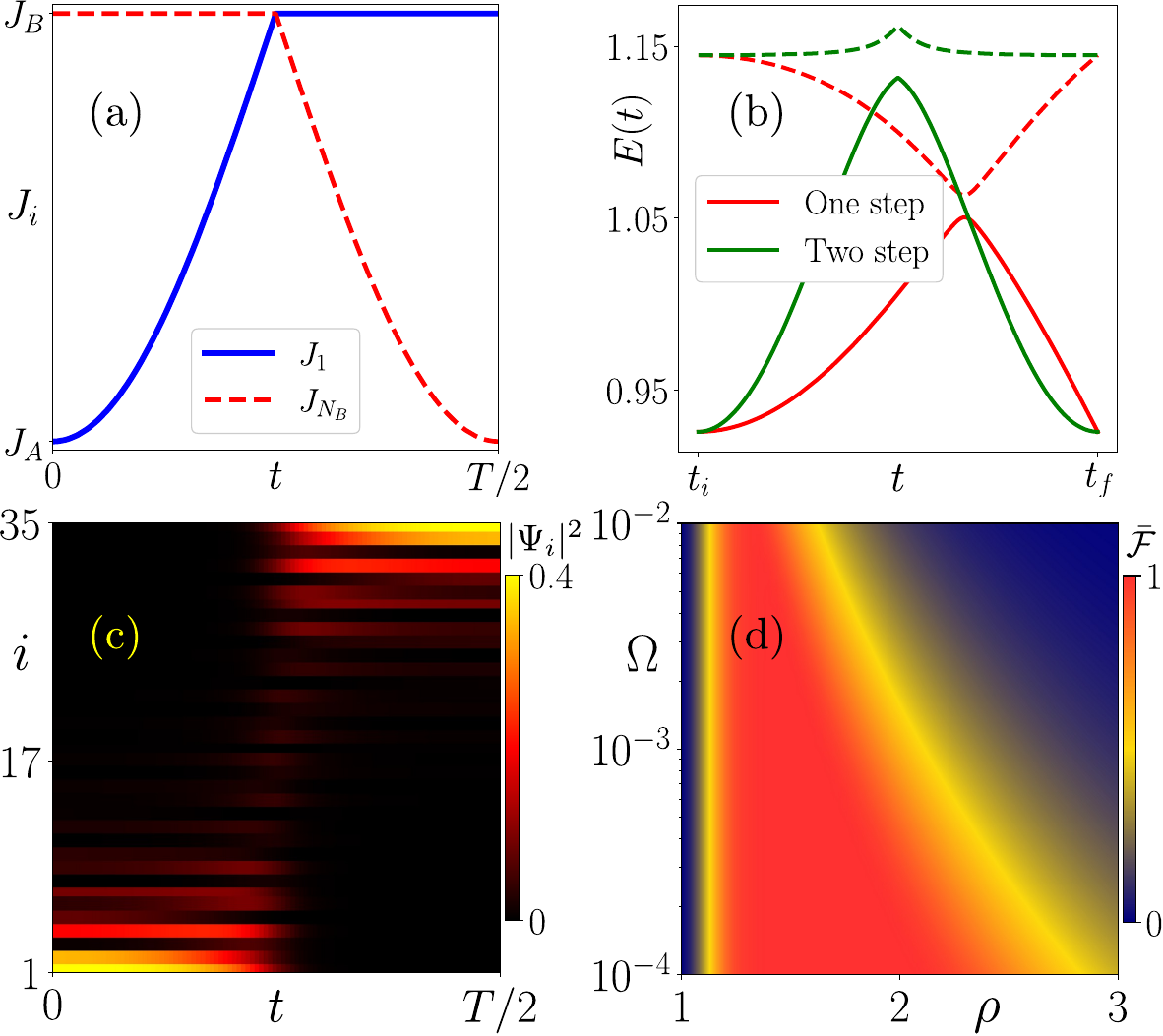}}
	\caption{(a) Dynamic change of hoppings $J_1$ and $J_{N_B}$ inside the unit cell of $C_8$ FC as a function of time in the two-step protocol, Eq.~\eqref{Eq:TwoStepHoppingChange}. (b) Comparison of the excitation gap between the winding state and the next excited state in the one-step, Eq.~\eqref{eq:J1JNB} (red) and two-step, Eq.~\eqref{Eq:TwoStepHoppingChange} (green) protocols. (c) Pumping of the winding state for the two-step protocol. (d) Weighted fidelity $\bar{\mathcal{F}}$ as a function of hopping ratio $\rho$ and rate $\Omega$ in the two-step protocol. The frequency axis is on a logarithmic scale. Here, $L=F_8+1=35$, $\rho=1.5$ and $\Omega=10^{-4}$ for (a-c).
	}
	\label{Fig:TwoStep}
\end{figure}
%~~~~~~~~~~~~~~~~~~~~~~~~~~~~~~~~~~~~~~~~~~~~~~~~~~~~~~~~~
%~~~~~~~~~~~~~~~~~~~~~~~~~~~~~~~~~~~~~~~~~~~~~~~~~~~~~~~~~

In the main text, we discuss some of the results of a two-step transfer protocol. In particular, as shown in Fig.~\ref{Fig:PhaseDiagram}(b) in the main text, the two-step protocol has an advantage of a larger region of high fidelity compared to the one-step protocol given by Eqs.~\eqref{Eq:driveprotcol},\eqref{eq:J1JNB}. Here, we discuss the details of the two-step protocol and some results associated with it.

In terms of the change in the outermost hoppings $J_1$ and $J_{N_B}$, the two-step protocol reads as
\begin{align}
    J_1 (t)=&J_B -(J_B-J_A)  \cos \Omega t \ , \quad t \in [0,T/4] , \non \\
    J_{N_B} (t)=&J_B +(J_B-J_A)  \cos \Omega t \ , \quad t \in [T/4,T/2].
    \label{Eq:TwoStepHoppingChange}
\end{align}
In Fig.~\ref{Fig:TwoStep}(a) we plot the time-dependence of these two hoppings, with the hopping $J_1$ (blue) first switched from $J_A$ to $J_B$ and only once that is completed we start switching the other hopping $J_{N_B}$ (red) from $J_B$ to $J_A$, to be compared to the one-step protocol illustrated in the inset of Fig.~\ref{Fig:QubitCoupling}(a) in the main text. In Fig.~\ref{Fig:TwoStep}(b), we then plot the evolution of the eigenvalues of the winding state (solid) and the next excited state (dashed) for one-step (red) and two-step (green) protocols. Since the final times for the one- and two-step protocols are different, we here, for plotting purposes, scale them such that $t_f$ is the same for both protocols. We find that the excitation gap isolating the winding state from the next excited state for the two-step protocol is slightly larger compared to the one-step protocol. Although this difference is small, it has a notable effect on the FQP operations as explained below. We also note that at the minimum gap, the evolution of the winding state and the excited state display a clear kink. This kink appears at the time instance when we interchange which hopping is being changed. We can therefore attribute the kink-like feature in the excitation gap to the fact that the changes in the hoppings also display kinks.

In Fig.~\ref{Fig:TwoStep}(c), we show the pumping of the winding state employing the two-step protocol. In comparison to the one-step protocol, see Fig.~\ref{Fig:Timedyn}(c) in the main text, the transition period when the winding state lives on both ends of the system is longer. In Fig.~\ref{Fig:TwoStep}(d), we depict the phase diagram in terms of the weighted fidelity $\bar{\mathcal{F}}$ (colorbar) as a function of the hopping ratio $\rho$ and rate $\Omega$ for the two-step protocol. The region with $\bar{\mathcal{F}} \simeq 1$ is more extended for the two-step protocol compared to the one-step protocol displayed in Fig.~\ref{Fig:PhaseDiagram}(a) in the main text. Thus, these results supplement the results shown in Figs.~\ref{Fig:Timedyn} and \ref{Fig:PhaseDiagram} in the main text and demonstrate how a different transfer protocol can lead to a wider region of parameters for successful FQP operation.

%======================================================
\section{Onsite Fibonacci chain} \label{App:Onsitemodel}
%======================================================

%~~~~~~~~~~~~~~~~~~~~~~~~~~~~~~~~~~~~~~~~~~~~~~~~~~~~~~~~~
%~~~~~~~~~~~~~~~~~~~~~~~~~~~~~~~~~~~~~~~~~~~~~~~~~~~~~~~~~
\begin{figure}
	\centering
	\subfigure{\includegraphics[width=0.48\textwidth]{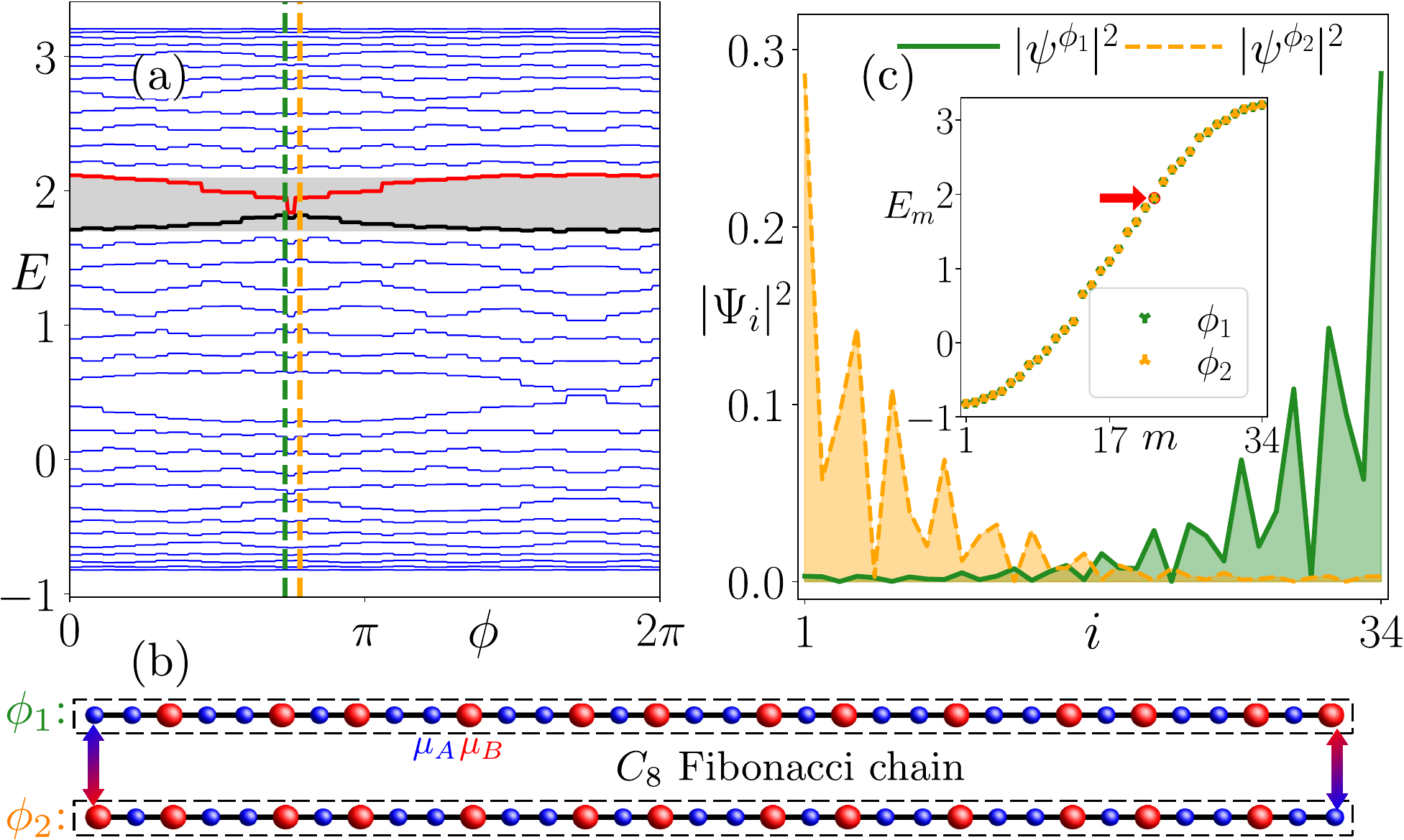}}
	\caption{(a) Eigenvalues $E $ as a function of phason angle $\phi$. Red curve represents a winding state in one of the quasicrystal gaps (gray shaded). (b) Flipping between high (red) and low (blue) onsite chemical potentials, indicated by the arrows when changing phason angles from $\phi_1$ (green) to $\phi_2$ (orange), also marked in (a). (c) Eigenstate of the winding state (red) along the chain at $\phi_1$ (green) and $\phi_2$ (orange). Inset shows eigenvalue spectrum $E_m$ as a function of the state index $m$ for phason angles $\phi_1$ (green) and $\phi_2$ (orange), with red dot marking the winding state. Here we use the $C_8$ FC with $L=F_8=34$ sites and $\tilde{\rho}=1.5$, $\phi_1=0.73 \pi$, and $\phi_2=0.78 \pi$. 
	}
	\label{Fig:Onsite}
\end{figure}
%~~~~~~~~~~~~~~~~~~~~~~~~~~~~~~~~~~~~~~~~~~~~~~~~~~~~~~~~~
%~~~~~~~~~~~~~~~~~~~~~~~~~~~~~~~~~~~~~~~~~~~~~~~~~~~~~~~~~

In the main text, we discuss results based on a hopping FC model. However, we can also build an FC employing an onsite FC model. Here, show that a FQP can also be constructed using winding states of an onsite FC.

In the onsite FC, the onsite chemical potential $\mu_i$ varies according to the Fibonacci sequence, while the hopping $J$ remains constant.  We can write the Hamiltonian for an onsite FC as~\cite{JagannathanRMP2021}
\begin{align}
    H= \sum_i \mu_i c_i^\dagger c_i +  J \sum_i c_i^\dagger c_{i+1} + {\rm H.c.} \ ,
    \label{Eq:HamOnsite}
\end{align}
where, the site-dependent $\mu_i$ take values as low ($\mu_A$) and high ($\mu_B$) chemical potentials, with $\left\{\mu_i\right \}$ belonging to the $n$-th Fibonacci approximant $\left\{ C_n \right\}$, such that $C_0=\mu_B$, $C_1=\mu_A$, $C_2=\mu_A \mu_B$, $C_3=\mu_A\mu_B\mu_A$, $C_4=\mu_A\mu_B\mu_A\mu_A \mu_B$, and so on~\cite{JagannathanRMP2021}. Thus, the total number of sites in $C_n$ equals the Fibonacci number $F_n=F_{n-1} + F_{n-2}$ $\forall n \geq 2$, with $F_0=1$, $F_1=1$. Note that the number of bonds in a unit cell is $N_B=F_n-1$ here, while the hopping model has $N_B=F_n$ number of bonds. Similar to the FC hopping model, we can construct longer onsite FCs by either considering larger approximants $C_n$, or repeating the approximant $C_n$ $l$ times, such that the total number of lattice sites of the chain is given as $ L=lF_n$. We also again employ open boundary condition~(OBC) and consider a dimensionless parameter $\tilde{\rho}=\mu_B/\mu_A$, while setting $\mu_A=J=1$ throughout this section and expressing all variables with the dimensions of energy in units of the hopping amplitude $J$.

We consider the $C_8$ FC with $L=F_8=34$ lattice sites and demonstrate the eigenvalue spectra as a function of the phason angle $\phi$ in Fig.~\ref{Fig:Onsite}(a). The red energy in  Fig.~\ref{Fig:Onsite}(a) marks a winding states inside a quasicrystal gap shaded by gray. In this quasicrystal gap, the winding state `winds' once, and the gap has gap label $\mathcal{C}=1$. Next, we focus on two specific values of the phason angle $\phi_1$ and $\phi_2$, denoted by the dashed green and orange lines in Fig.~\ref{Fig:Onsite}(a). We choose these $\phi_{1,2}$ such that the change of the phason angle from $\phi_1$ to $\phi_2$ amounts to only flipping the two outermost chemical potentials in a unit cell of the $C_8$ FC, as denoted by the arrows in Fig.~\ref{Fig:Onsite}(b). This is analogous to the setup in Fig.~\ref{Fig:FQCbackground}(b) in the main text. We denote these two chemical potentials as $\mu_1$ and $\mu_{N_B+1}$. We plot the eigenvalue spectrum for $\phi_1$ (green) and $\phi_2$ (orange) in the inset of Fig.~\ref{Fig:Onsite}(c) and see that they perfectly overlap, with with winding state marked in red appearing around $E_m\simeq 2.0$. Then, in Fig.~\ref{Fig:Onsite}(c), we investigate the spatial extent of the winding state for the phason angles $\phi_1$ (green) and $\phi_2$ (orange) and find that it completely switches between the two ends of the chain with the change of phason angle. These results are thus analogous to Fig.~\ref{Fig:FQCbackground} in the main text and demonstrate that winding states in the onsite FC can also be employed for FQP operation.

%~~~~~~~~~~~~~~~~~~~~~~~~~~~~~~~~~~~~~~~~~~~~~~~~~~~~~~~~~
%~~~~~~~~~~~~~~~~~~~~~~~~~~~~~~~~~~~~~~~~~~~~~~~~~~~~~~~~~
\begin{figure}
	\centering
	\subfigure{\includegraphics[width=0.48\textwidth]{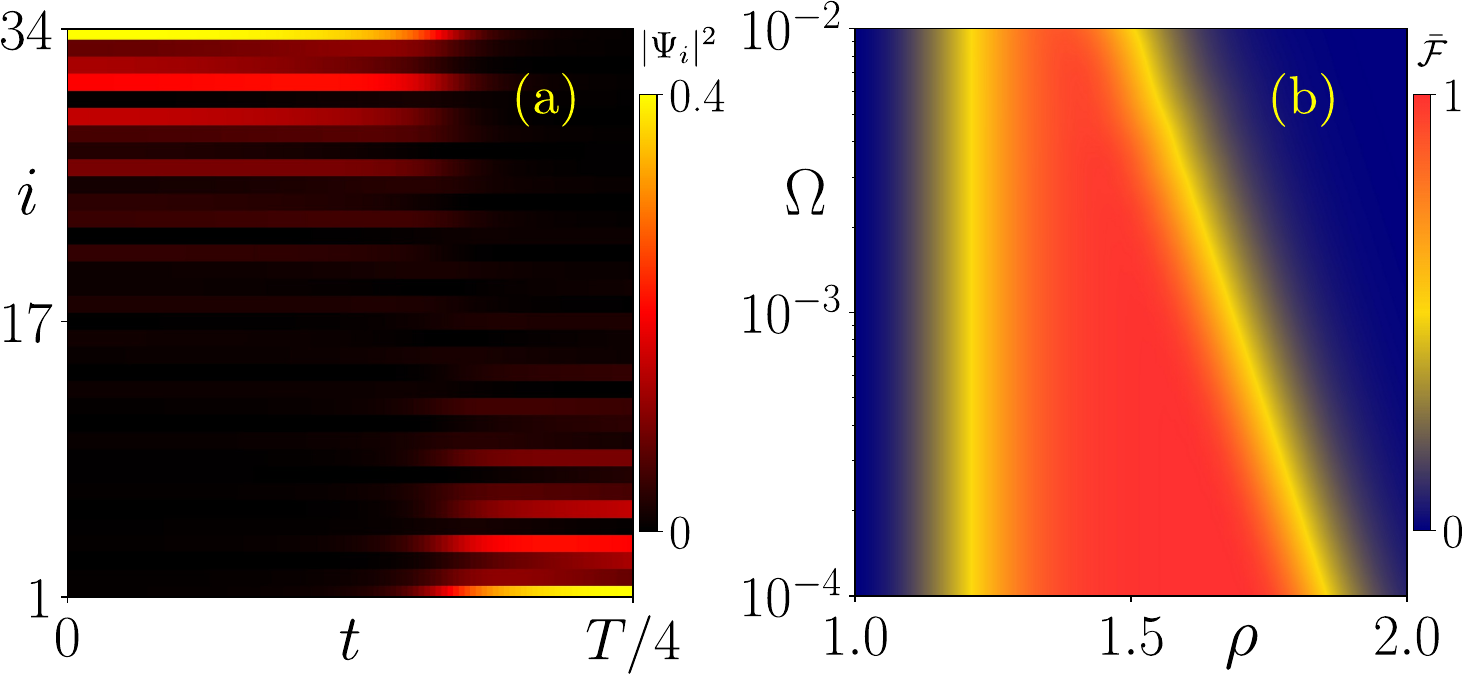}}
	\caption{(a) Evolution of the winding state as a function of time $t$ and spatial position $i$ for the $C_8$ FC in Fig.~\ref{Fig:Onsite}. Here $\Omega=10^{-4}$ and $\rho=1.5$. (b) Weighted fidelity $\bar{\mathcal{F}}$ as a function of hopping ratio $\tilde{\rho}$ and rate $\Omega$. The frequency axis is on a logarithmic scale.
	}
	\label{Fig:OnsiteDynamics}
\end{figure}
%~~~~~~~~~~~~~~~~~~~~~~~~~~~~~~~~~~~~~~~~~~~~~~~~~~~~~~~~~
%~~~~~~~~~~~~~~~~~~~~~~~~~~~~~~~~~~~~~~~~~~~~~~~~~~~~~~~~~

For completeness, we explicitly demonstrate FQP operations for the onsite FC. We consider a similar one-step transfer protocol as for the hopping FC model in the main text, i.e.~Eq.~\eqref{Eq:driveprotcol}. The transfer protocol thus reads
\begin{align}
    V(t)= \left\{ \mu_i^{\phi_1} \right\} + \left( \left\{ \mu_i^{\phi_2} \right\}- \left\{ \mu_i^{\phi_1} \right\} \right) (1-\cos \Omega t) \ , \quad t \in [0,T/4] ,
    \label{Eq:Onsitedrive}
\end{align}
where $\left\{ \mu_i^{\phi_i} \right\}$ represents the chemical potential sequence corresponding to phason angle $\phi_i$ and $\Omega$ indicates the rate. Note that there is only a change of two chemical potentials in a unit cell in $V(t)$ as shown in Fig.~\ref{Fig:Onsite}(b), while others remain the same, similar to the hopping FC transfer protocol employed in the main text. More explicitly, in terms of the chemical potential $\mu_{1,N_B+1}$, the transfer protocol reads as
\begin{align}
    \mu_1 (t)=&\mu_B -(\mu_B-\mu_A)  \cos \Omega t \ , \quad t \in [0,T/4] , \non \\
    \mu_{N_B+1} (t)=&\mu_A +(\mu_B-\mu_A)  \cos \Omega t \ , \quad t \in [0,T/4] \ .
\end{align}
Using the one-step transfer protocol and solving for the full time-dependence as outlined in the Appendix~\ref{App:Method}, we demonstrate the pumping in the onsite FQP. In the main text, we show pumping of a state from left to right. Here, we display an opposite case, where we transfer a state from the right end to the left. We depict the time-resolved state transfer in Fig.~\ref{Fig:OnsiteDynamics}(a), showing a winding state which is initially localized at the right end of the system being transferred to the other left end. We also compute the weighted fidelity $\bar{\mathcal{F}}$ and plot it as a function of $\tilde{\rho}$ and $\Omega$ in Fig.~\ref{Fig:OnsiteDynamics}(b), thus showing complementary information to Fig.~\ref{Fig:PhaseDiagram}(a) in the main text.
These results demonstrate that we can equally well use an onsite FC to perform quantum state transfer using the winding state.

\begin{onecolumngrid}
%======================================================
\section{State transfer and entanglement generation} \label{App:Entanglement}
%======================================================
In the main text, we discuss state switching and entanglement generation employing the FQP and using the initial state $\ket{101}$ and the one-step protocol. Here, we provide complementary results using other initial states and protocols.
First, we discuss results for different initial states. In Fig.~\ref{Fig:OtherInitialStateFig4}(a) and (b), we show the coefficients $\lvert \alpha_{1,2,3}(t) \rvert^2$ of the time-evolved state for the initial many-body states $\ket{110}$ and $\ket{011}$, respectively. For the initial state $\ket{110}$, we do not observe any change in the time-evolved state. This is expected as both qubits are occupied initially, so there is no change in the qubit state from FQP operations. However, for the initial state $\ket{011}$, we see an analogous behavior to the state $\ket{101}$ in Fig.~\ref{Fig:QubitCoupling}(b) in the main text. 
%~~~~~~~~~~~~~~~~~~~~~~~~~~~~~~~~~~~~~~~~~~~~~~~~~~~~~~~~~
%~~~~~~~~~~~~~~~~~~~~~~~~~~~~~~~~~~~~~~~~~~~~~~~~~~~~~~~~~
\begin{figure}
	\centering
	\subfigure{\includegraphics[width=0.5\textwidth]{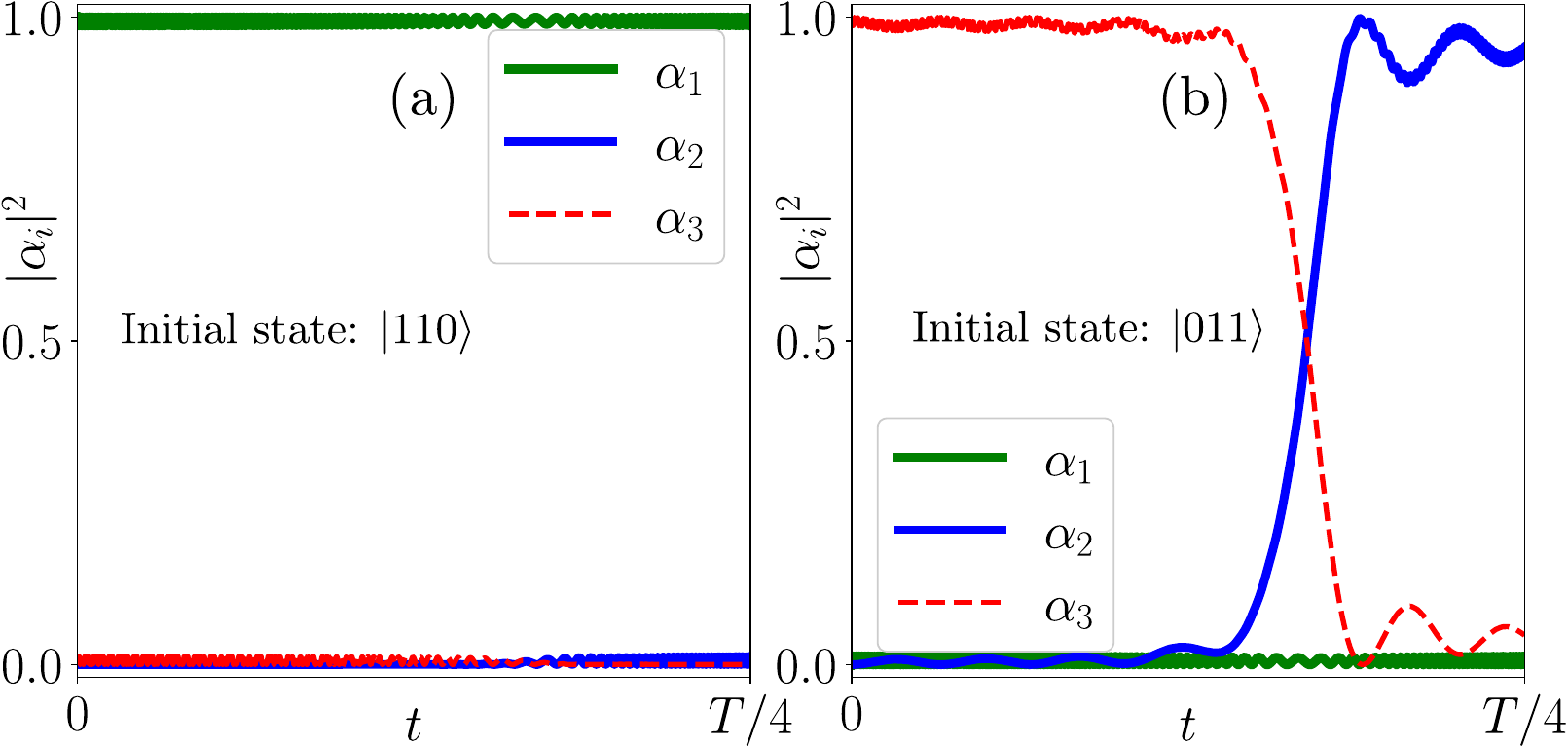}}
	\caption{(a) Amplitudes $\lvert \alpha_{1,2,3}(t) \rvert^2$ of the time-evolved many-body state for an initial state (a) $\ket{110}$ and (b) $\ket{011}$ as a function of time. The figures reproduce Fig.~\ref{Fig:QubitCoupling}(b) from the main text but with other initial states, while all other parameters are the same, except $g=0.1$ to reduce the oscillation of the time-evolved state.
	}
	\label{Fig:OtherInitialStateFig4}
\end{figure}
%~~~~~~~~~~~~~~~~~~~~~~~~~~~~~~~~~~~~~~~~~~~~~~~~~~~~~~~~~
%~~~~~~~~~~~~~~~~~~~~~~~~~~~~~~~~~~~~~~~~~~~~~~~~~~~~~~~~~

In the main text, we discuss how the two-step protocol can be employed to expand the parameter space for high fidelity of the FQP. Here, we complement the results also for coupling the FQP to the two qubits. In particular, we demonstrate state transfer operation between two qubits when employing the two-step protocol in Fig.~\ref{Fig:TwoStepEntanglement}. We first show the transfer of the winding state in the FQP in Fig.~\ref{Fig:TwoStepEntanglement}(a). In Fig.~\ref{Fig:TwoStepEntanglement}(b), we then show how the coefficients $\lvert \alpha_{1,2,3}(t) \rvert^2$ of the time-evolved many-body state behave for an initial state $\ket{101}$ when transferring the winding state. As expected, we observe a similar behavior to Fig.~\ref{Fig:QubitCoupling}(b) in the main text, with the qubit states getting flipped as we complete the transfer protocol. This behavior is reflected in the concurrence as well as seen in Fig.~\ref{Fig:TwoStepEntanglement}(c). The concurrence exhibits a value close to $1$ when the qubit states are being flipped but we do not retain a high concurrence at the end of the transfer protocol. This behavior is also observed when we plot the overlap of the time-evolved state with the Bell states ($P_{\rm B}^{1,2}$) in the inset of Fig.~\ref{Fig:TwoStepEntanglement}(c). We obtain the Bell state for a short instance of time when the concurrence is $1$. Thus, we observe a fully analogous scenario for the two-step protocol as that of the one-step protocol in Fig.~\ref{Fig:QubitCoupling}(c) in the main text. 

%~~~~~~~~~~~~~~~~~~~~~~~~~~~~~~~~~~~~~~~~~~~~~~~~~~~~~~~~~
%~~~~~~~~~~~~~~~~~~~~~~~~~~~~~~~~~~~~~~~~~~~~~~~~~~~~~~~~~
\begin{figure}
	\centering
	\subfigure{\includegraphics[width=0.7\textwidth]{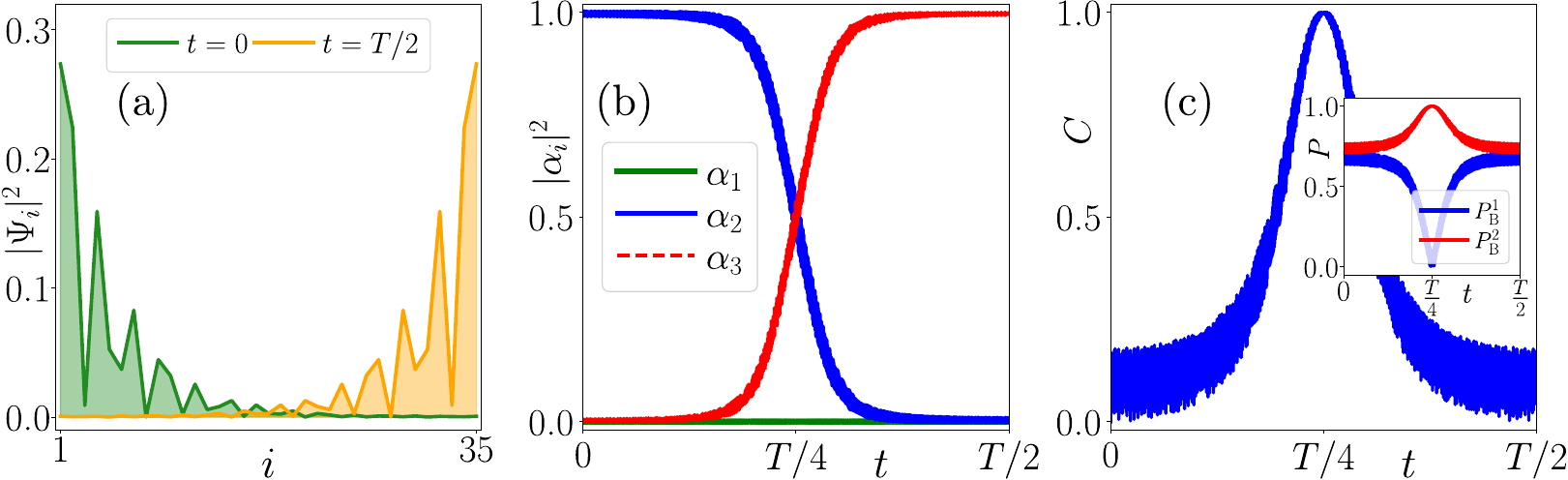}}
	\caption{(a) Spatially resolved amplitude of the winding state at initial time $t=0$~(green) and final time $t=t_f$~(orange) using the two-step transfer protocol in Fig.~\ref{Fig:TwoStep}. (b) Amplitudes $\lvert \alpha_{1,2,3}(t) \rvert^2$ of the time-evolved many-body state for an initial state $\ket{101}$ as a function of time. (c) Concurrence $C$ as a function of time for the same initial state as in (b). Inset shows the probability of obtaining the two Bell states. The figures reproduce Fig.~\ref{Fig:QubitCoupling}(a-c) from the main text but with the two-step protocol, while all other parameters are the same.
	}
	\label{Fig:TwoStepEntanglement}
\end{figure}
%~~~~~~~~~~~~~~~~~~~~~~~~~~~~~~~~~~~~~~~~~~~~~~~~~~~~~~~~~
%~~~~~~~~~~~~~~~~~~~~~~~~~~~~~~~~~~~~~~~~~~~~~~~~~~~~~~~~~

%~~~~~~~~~~~~~~~~~~~~~~~~~~~~~~~~~~~~~~~~~~~~~~~~~~~~~~~~~
%~~~~~~~~~~~~~~~~~~~~~~~~~~~~~~~~~~~~~~~~~~~~~~~~~~~~~~~~~
\begin{figure}[h]
	\centering
	\subfigure{\includegraphics[width=0.7\textwidth]{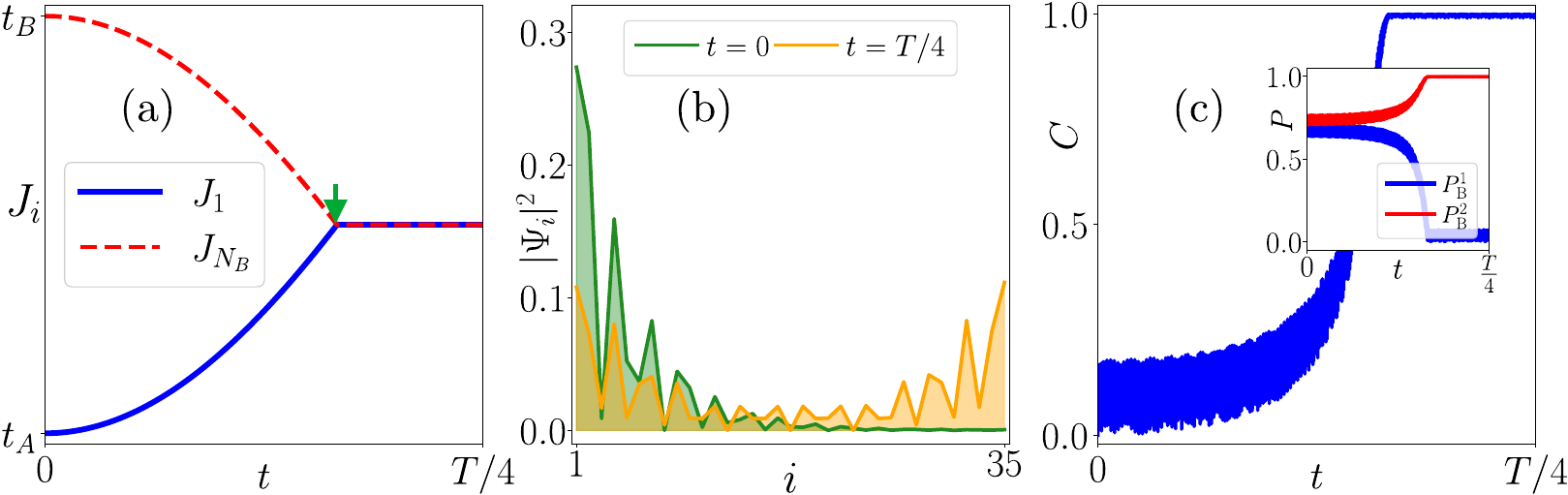}}
	\caption{(a) Modified one-step transfer protocol as a function of time $t$ with green arrow marking the point when we turn off the protocol. (b) Spatially resolved amplitude of the winding state at initial time $t=0$~(green) and final time $t=t_f$~(orange) using the transfer protocol in (a). (c) Concurrence $C$ for the initial state $\ket{101}$ as a function of time. Inset shows the probabilities of obtaining the two Bell states. The figures reproduce Fig.~\ref{Fig:QubitCoupling}(a,c) from the main text but with the turned off one-step protocol in (a), while all other parameters are the same.
	}
	\label{Fig:OnestepSwitching}
\end{figure}
%~~~~~~~~~~~~~~~~~~~~~~~~~~~~~~~~~~~~~~~~~~~~~~~~~~~~~~~~~
%~~~~~~~~~~~~~~~~~~~~~~~~~~~~~~~~~~~~~~~~~~~~~~~~~~~~~~~~~

Finally, in the main text in Fig.~\ref{Fig:QubitCoupling}(c), we observe near max $C \simeq 1$ when the winding state is switched from one side to the other of the system. However, we can then not retain this high concurrence $C$ at the final time. Here, we show that by simply turning off the one-step protocol, we can also maintain the high concurrence at the final time $t_f$. Thus, we stop the change of the hoppings at the point marked by the green arrow in Fig.~\ref{Fig:OnestepSwitching}(a). When we stop our parameter-changing protocol at this point, the winding state does not get completely transferred to the other end of the system, rather, it is now localized at both ends of the chain in a superposition state, as shown in Fig.~\ref{Fig:OnestepSwitching}(b). At the same time, by stopping the transfer protocol here, we retain the maximum $C \simeq 1$ until the final time, as shown in Fig.~\ref{Fig:OnestepSwitching}(c). We also show the overlap of the time-evolved state with the Bell state in the inset of Fig.~\ref{Fig:OnestepSwitching}(c), observing that we obtain one of the Bell states.

\end{onecolumngrid}

\begin{twocolumngrid}
\end{twocolumngrid}
\bibliographystyle{apsrev4-2mod}
\bibliography{bibfile.bib}

%============End of MAIN PAPER=============

\end{document}